# Buoy observation of high frequency ocean wave energy: accuracy, consistency, and concerns for predictive applications

W. Erick Rogers

Naval Research Laboratory, Stennis Space Center, MS, USA

Jim Thomson

Applied Physics Laboratory, University of Washington, Seattle WA, USA

Corresponding author: W. Erick Rogers (w.e.rogers.civ@us.navy.mil)

## Executive Summary

Observational data from buoys are of primary importance during the development, calibration, and evaluation of ocean wave models, and these data are also used to make real-time corrections to operational models via data assimilation. By association, systematic inaccuracies in any buoy data are equally important, and thus when two buoy types provide systematically inconsistent information, this is a concern for anyone using an ocean wave model. This report is concerned with the accuracy of the high frequency portion of the ocean wave spectrum commonly observable by buoys, roughly 0.2 to 0.6 Hz. We evaluate four buoy types (two moored, two drifting) using two quantitative measures. The first involves comparing each type with a co-located ocean wave model. The second method involves evaluation of high frequency energy level as a function of wind speed. Both evaluation methods suggest that the Datawell Waverider (DWR) buoys have a strong tendency to report higher energy levels than the other three buoy types. A possible explanation is the Doppler shift of the drifting buoys and a damped response of the larger moored buoys. We evaluate high frequency energy level using three different metrics (mean square slope, energy in a band of high frequencies, and spectral density at a single, specific band, 0.4 Hz), and the conclusions are found to be insensitive to the parameter used.

## 1. Introduction

In situ observations of ocean wave spectra are needed for ground-truthing of wave spectra that are observed by satellite and predicted by numerical wave models. In the deeper oceans—say depths greater than 30 m—the most common in situ method is to compute the spectrum from the motion of a floating buoy, e.g., Longuet-Higgins et al. (1963), Steele et al. (1985)[1]. Traditionally, these buoy networks are composed of moored buoys, e.g., Hemsley (1996), but more recently, large networks of drifting buoys have been deployed, e.g., Houghton et al. (2021). The designs of wave buoys are highly varied, both in hulls and motion-sensing instrumentation. Generally speaking, they measure bulk energy[2] or significant wave height $H_s$ with high skill, e.g., O'Reilly et al. (1996) and Jensen et al. (2021), though problems are sometimes noted under extreme conditions, e.g., Bender et al. (2010). Observations of directional distribution are less reliable with some buoys, e.g., O'Reilly et al. (1996), as are observations of the energy level at low frequencies, where the signal-to-noise ratio can present problems, and high frequencies, where buoy response functions—and potentially, other issues—make accurate measurement more difficult, e.g., Section 4.3 of Collins et al. (2024). This report is concerned with the high frequencies, which have particular importance to some Navy applications.

The accuracy of buoy-derived ocean wave observations is relevant to three aspects of ocean wave modeling: evaluation, calibration, and data assimilation. In the context of low-order parameters such as significant wave height, there are many examples of each aspect, but in the context of wave spectra, this list is sharply reduced. Some examples are Rogers and van Vledder (2013) (evaluation), Ardhuin et al. (2010) (evaluation and calibration), and Voorips et al. (1997) (data assimilation). Combining two themes raised thus far, Houghton et al. (2022) describes spectral data assimilation using large drifting buoy networks. Before the buoy data can be used for any of these purposes, we need to answer questions such as: Can we use the data as-is, or can corrections be made to offset systemic bias? What is the uncertainty in the data, and how does this depend on frequency, energy level (signal strength), or the type of parameter[3]?

When multiple instruments produce multiple suggestions regarding the need for model recalibration, this is especially troublesome. The first step in addressing this is to quantify these inconsistencies. In this report, we do this using two methods of comparing the high frequency energy spectra from dissimilar wave buoys which are, for the most part, not co-located. The first method is to compare both to an intermediate "data" source: the ocean wave model. The second method is to evaluate the mean response to variations in wind speed. These methods are described in Section 2. The first method (involving the use of a numerical wave model) is applied in Section 3. The second method (based on mean trends with wind speed) is applied in Section 4. In Section 5, we evaluate one of the wave buoy types using visual inspection of spectral slope; this is a more qualitative and subjective method than those applied in Sections 3 and 4, but provides useful, independent insights. In Section 6, we discuss possible reasons for discrepancies observed in Sections 3 and 4. In Section 7, we summarize the results and make recommendations for future work.

[1] The wave spectrum can also be estimated from a pressure gage or via acoustic surface tracking, but these methods are less common in the deep ocean, and the instruments must not be too far below the sea surface.
[2] 'Bulk energy' is used here to describe the energy computed from integration of a large frequency range, including frequencies near the peak, i.e., the 'energy-containing' region of the spectrum.
[3] As an example, take the directional distribution of energy at each frequency. The mean direction may be easiest to estimate, and perhaps direction spread is less reliable, and skewness and kurtosis of the distribution are yet more difficult (e.g., O'Reilly et al. 1996).

## 2. Methods

### 2.1. Parameters used

The moments of the one-dimensional sea surface elevation spectrum $E(f)$ are computed as $m_n = \int_{f_1}^{f_2} E(f) f^n df$, where $f_1$ and $f_2$ are the lower and upper bounds of integration over frequency $f$. Frequency has units Hz and spectral density $E(f)$ has units $m^2$/Hz. The energy in a frequency band of ocean waves is computed as $m_{0B} = \int_{f_1}^{f_2} E(f) df$. This energy (or more precisely, the contribution to the total variance of the sea surface elevation by that band of frequencies) is given in units $m^2$ or $cm^2$. For example, O'Reilly et al. (1996) use $f_1 = 0.06$ Hz and $f_2 = 0.14$ Hz to evaluate locally dominant swell. Here "B" denotes "band" or "between".

Rather than using energy, we convert to an equivalent waveheight (units m) as was done by Rogers (2002). The traditional calculation for wave height is $H_{m0} = 4\sqrt{m_0}$, and so the band-wise calculation is simply $H_{m0B} = 4\sqrt{m_{0B}}$. Customarily, we determine the full range of overlap in frequencies for model and observation, and break into four bands, e.g., Rogers et al. (2005) and Rogers (2023) use bands delineated by $f$, and Rogers and Wang (2007) use four bands delineated by $f/f_p$, where $f_p$ is the time-dependent peak frequency. In the current study, however, our attention is on the fourth of four bands $H_{m0B,4}$, i.e., the high frequency band.

The Coastal Data Information Program (CDIP) of UCSD/SIO also reports the $H_{m0B}$ parameter, referring to it as "band height", split into nine bands. Liu et al. (2019) and Collins et al. (2024) used four bands, and Liu et al. (2020) used three.

For sake of consistency with earlier comparisons such as Rogers (2017), we also employ the parameter $m_4$, which, when integrated over all wave frequencies, is proportional to the mean square slope (mss) of the sea surface. When integrated over a limited band of frequencies, e.g., "up to $f_{max} = 0.5$ Hz", it is sometimes called a "low-pass $m_4$" or converted to a "low-pass mss" (Chen et al. 2016). This terminology acknowledges the dominant influence of higher frequencies on $m_4$, such that the lower bound $f_1$ is often inconsequential, for example, $m_4 = \int_{0.04}^{0.5} E(f) f^4 df \approx \int_{0}^{0.5} E(f) f^4 df$.

The two parameters have an important difference. $H_{m0B,4}$ often excludes the spectral peak, i.e., it is (often) quantifying the energy in the spectral tail rather than the "energy-containing region" of the spectrum. This is especially true when both $f_1$ and the local wind speed are larger, e.g., $f_1$>0.25 Hz and $U_{10}$>10 m/s. By contrast, the region around the spectral peak usually has a significant impact on $m_4$.

### 2.2. Observational datasets used

We employ spectra from several buoy observational datasets:

1) The Ocean Station Papa (OSP) buoy of University of Washington's Applied Physics Laboratory (UW/APL), which is CDIP buoy #166. Its position is shown in Figure 1. We primarily use the period 16 May 2022 to 17 May 2023, but data from other time periods are used in Figure 9 and Figure 15. Data were downloaded from the CDIP website.
2) Drifting mini-wave-buoys deployed by University of California, San Diego (UCSD) Scripps Institute of Oceanography (SIO) Coastal Observing Research and Development Center (CORDC), provided by E. Terrill, S. Merrifield, and others. The time period of the data is 24 November 2016 to 11 April 2017 (thus, 4.6 months duration). The drifter positions are shown in Figure 1.
3) Drifting Spotter buoys deployed by Sofar Oceans. These buoys measure wave spectra over much of the world's oceans. We use data from the period September 2021 to May 2022, provided by P. Smit of Sofar. These drifter positions are shown in Figure 2.
4) 13 UCSD/SIO/CDIP buoys including the OSP buoy, for up to 32 months duration per buoy, depending on outages, May 2022 to December 2024. Data were downloaded from the CDIP website. These 13 mooring positions are shown in Figure 2.
5) 8 National Oceanic and Atmospheric Administration (NOAA) National Data Buoy Center (NDBC) buoys, for up to 12 months duration per buoy, depending on outages, January to December 2024. Data were downloaded from the NDBC website. These 8 mooring positions are shown in Figure 2.

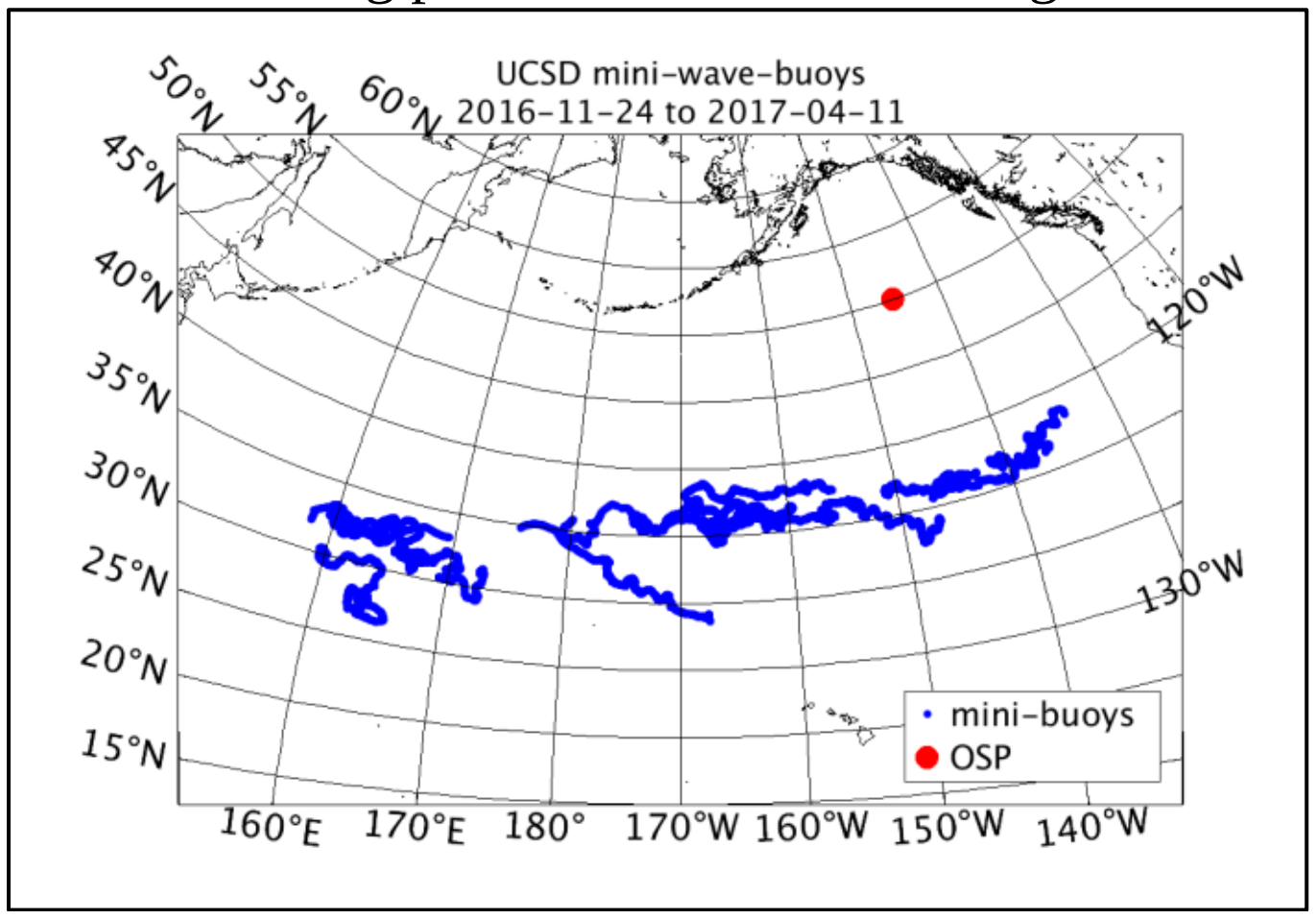

Figure 1. Buoy positions. Blue tracks: UCSD/SIO/CORDC buoys, 24 November 2016 to 11 April 2017. Red dot: UW/APL buoy at OSP (CDIP 166).

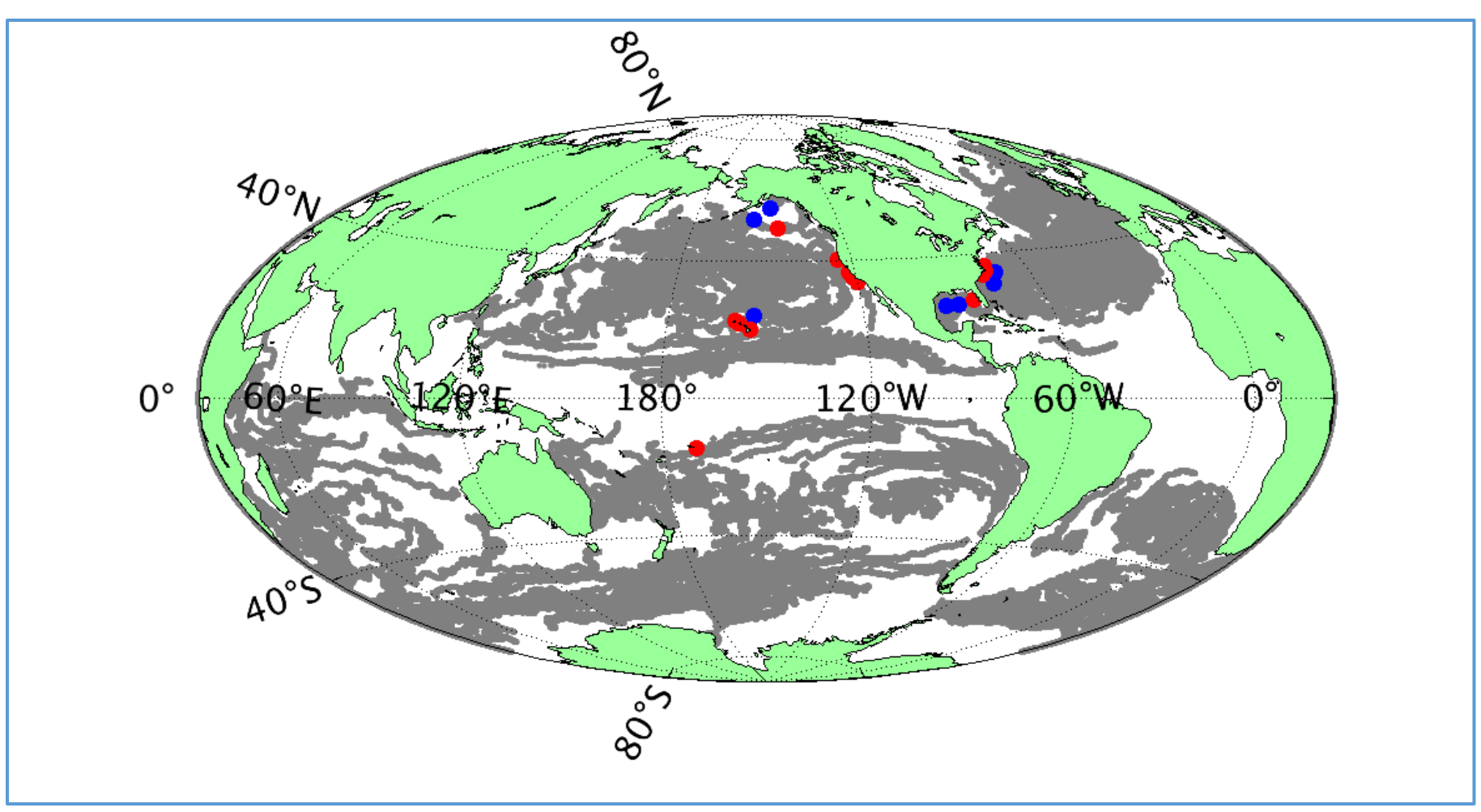

Figure 2. Buoy positions. The color of the markers indicates the type of buoy: gray for Sofar, red for CDIP, and blue for NDBC.

## 2.3. Model-based evaluation (method description)

Our first method of evaluation is to compare different buoy datasets to a numerical wave model, which serves as an intermediary. Consistency, or lack thereof, in the model-buoy comparisons tells us whether the buoys are consistent.

The primary shortcoming of this method (i.e., source of uncertainty) is due to the differences in geography between the co-locations. Insofar as the model may have bias that is geographically heterogeneous, this will add error to the comparison. Temporal variability may also have an effect, insofar as model bias depends on seasonal[4] or year-to-year (e.g., ENSO) variability.

### *2.3.1. Model description*

The wave model used in this study WAVEWATCH III® (WW3, Tolman 1991, WW3DG 2019). This is a phase-averaged model for which the prognostic variable is wave action spectral density, which is the wave energy spectral density divided by the circular wave frequency: $N = E/\sigma$, where $\sigma = 2\pi f = 2\pi/T$ ($T$ denoting wave period). The spectrum is a function of wavenumber or frequency ($k$ or $\sigma$), direction ($\theta$), space ($x$, $y$ or longitude, latitude), and time ($t$). The left-hand side of the radiative transfer equation includes terms for time rate of change and propagation in the four dimensions (kinematics), while the right-hand side includes source functions (dynamics):

$$\frac{\partial N}{\partial t} + \nabla \cdot \vec{c} N = \frac{S}{\sigma}$$

Here, $\vec{c}$ is a four-component vector describing the propagation velocities in $x$, $y$, $k$, and $\theta$. For example, in absence of currents, $c_x$ is the $x$-component of group velocity $C_g$. The sum of all source functions is denoted as $S$, and individual source functions are denoted with appropriate subscript: $S_{in}$, $S_{wc}$, $S_{nl4}$, and $S_{ice}$ being energy input from wind, dissipation by whitecapping, four-wave nonlinear interactions, and dissipation by sea ice, respectively. In deep water, without ice cover, the terms $S_{in}$, $S_{wc}$, and $S_{nl4}$ dominate $S$.

[4] Our shortest time series is 4.6 months duration, so it does not capture the seasonal variability.

Except where otherwise noted, the specific implementation of WW3 used in this study follows the description below.

The model version used is based on a development version of WW3, intermediate between 7.00 and 7.01, including changes by NRL that are not relevant to the present report.

Open water source terms and spectral grid settings used here are fairly typical of routine large-scale modeling using WW3. We use the "source term package" of Ardhuin et al. (2010) known as 'ST4', for $S_{in}$ and $S_{wc}$. In this package, swell dissipation (weak losses of energy not associated with breaking) is formally part of $S_{in}$. For $S_{nl4}$, we use the Discrete Interaction Approximation (DIA) of Hasselmann et al. (1985).

Wind forcing and ice concentration fields are taken from the ECMWF (European Centre for Medium-Range Weather Forecast) Reanalysis v5 (ERA5) reanalysis (Hersbach et al. 2020), at 1-hourly intervals and 1/4˚ geographic resolution. The wind forcing are in the form of 10-m neutral wind vectors[5]. Currents are not included in these hindcasts.

In a few cases (Sections 3.2 and 3.3), we present wave model results using the ERA5 forcing just described, and also an alternative source of forcing: U.S. Navy's operational global atmospheric model, NAVGEM (Navy Global Environmental Model, Hogan et al. 2014), with 10-m non-neutral wind vectors 3-hourly at 0.28° resolution and ice concentration analyses 12-hourly at 0.5° resolution[6]. These additional model runs serve two purposes: first, via comparison with the ERA5 runs, it quantifies the sensitivity to the forcing, and second, via comparison with runs performed by the author in 2017 using NAVGEM forcing, it quantifies the sensitivity to the change in model version and settings.

The global grid design is IRI-1/4. This is the "Irregular-Regular-Irregular" design (Rogers and Linzell 2018, Fan et al. 2021), with resolution of 1/4˚ at low latitudes and 18 km at latitudes higher than 50˚. We use an overall time step size of 1800 s, a propagation time step of 600 s. The source term time step is dynamically determined by WW3; in our implementation, it is not allowed to be less than 10 s.

The spectral grid includes 36 directional bins and 32 frequency bins (0.038 to 0.73 Hz, logarithmically spaced). The wind input source term of Ardhuin et al. (2010) requires specification of a parameter, $\beta_{max}$ which is used to compensate for the mean bias of the input wind fields, or lack thereof. We use $\beta_{max}$=1.43 for the hindcasts with ERA5 forcing and $\beta_{max}$=1.20 for the hindcasts with NAVGEM forcing.

The ST4 physics package also has a parameter denoted 'FXFM3' in the user settings and $f_{FM}$ in Ardhuin et al. (2010). This is particularly relevant to high frequencies, since it controls the frequency limit for the prognostic spectrum: after this limit, a diagnostic tail is attached. The relevant equation is equation (7) of Ardhuin et al. (2010): $f_c = f_{FM} f_m$, where $f_m$ is a mean frequency, $f_m = f_{m,0,1} = m_1/m_0$. See Ardhuin et al. (2010) for further context. We use the default value, and for version 7.00, that default is $f_{FM} = 2.5$.[7] When a larger $f_{FM}$ is used, this tends to increase high frequency energy levels for moderate to high wind speeds (Appendix A).

[5] 10-m neutral winds are generally preferred over standard 10-m winds for forcing wave models, but are not available from some atmospheric models.

[6] The resolution of the operational NAVGEM has changed over time. At time of writing (November 2025), the resolution of the wind and ice fields are both 0.18°.

### 2.4. Wind speed-based evaluation (method description)

The second method of evaluating the high frequency energy reported by buoys is to compare:

1) a wave parameter which is sensitive to the high frequency energy level

against

2) a wind speed parameter.

In a scatter plot presentation, (1) is the vertical axis and (2) is the horizontal axis. In this report, we use the following options for (1):

a) $H_{m0B}$ with $f_1$=0.29 Hz and $f_2$=0.58 Hz, and
b) $E(f)$ at $f$=0.4 Hz.

And we use these options for (2):

a) the 4-m wind speed $U_4$ measured by the NOAA/PMEL meteorological buoy at Ocean Station Papa,
b) the 10-m neutral wind speed $U_{10}$ provided by ERA5, and
c) $U_4$ estimated from $U_{10}$ with a simple approximation, $U_4 \approx 0.91 U_{10}$, where $U_{10}$ is (b). [This is from a "power law" approximation of the vertical wind profile, $(4/10)^{\frac{1}{10}} = 0.91$.]

This method of comparison suffers from the problems like those mentioned in our description of the model-based comparison: geographic and temporal variability of the unknown true (1) vs. (2) relation. It implicitly assumes that, given a sufficiently large dataset, the (1) vs. (2) relation regresses to a near-universal mean. In the case of 2b and 2c, it is also affected by geographic and temporal variability of errors in the wind speeds of ERA5, but we do not expect this to be a significant factor.

This comparison assumes a direct relation between wave energy (1) and wind speed (2), i.e., it assumes that (1) is wind sea. Thus, any swell energy in the comparison is a contaminant. In our plots of (1) vs. (2), where wind speed (2) is small, the value of wave energy (1) is more often contaminated by swell[8], and so that portion of the plot is less reliable. Thus, comparisons herein at low wind speeds are viewed with suspicion, and comparisons at very low wind speeds are omitted entirely.

Co-locations of hourly, 1/4˚ ERA5 wind speed with the buoys are created using nearest neighbor in time and bilinear interpolation in space.

---

[7] In Ardhuin et al. (2010) and in older versions of the code, the default FXFM3 is 9.9. In the NRL code repository, it changed to 2.5 between versions 5.13 and 5.15, sometime between September and November 2016.

[8] A more verbose explanation follows. Given a directional wave spectrum $E(f,\theta)$ in discretized form, with bins for each $f$ and $\theta$, such as is provided by a wave model, swell can be defined as any $(f,\theta)$ bin for which the wave phase velocity $c > U_p$, where $U_p = C_{mult} U_{10} \cos(\theta - \theta_w)$. Here, $C_{mult}$ is a multiplier (e.g., WW3DG (2019) uses a default $C_{mult} = 1.7$) and $\theta_w$ is the wind direction. This implies that, as wind speed decreases, a larger fraction of the spectrum is considered swell.

# 3. Model-based evaluation

## 3.1. Background

Before we show the first comparisons, we briefly explain how we got here. In April 2017, Dr. Eric Terrill (UCSD/SIO/CORDC) provided wave spectral observations (indicated in the mini-buoy tracks in Figure 1). The author (Rogers) ran a hindcast of WW3, co-located with the drifting buoy data, made comparisons, and reported this in a poster, Rogers (2017)[9]. The poster compared $m_4$ of WW3 against the UCSD/SIO/CORDC buoys, indicated a positive bias in the wave model, and stated "comparison with UW Datawell buoy (not included here) for the same simulation show similar bias in [$m_4$]". Though that plot, for the 2016/2017 period, was not presented, a plot was included in the poster, comparing two models, SWAN and WW3, against the Ocean Station Papa Datawell buoy (Figure 1) for a different time period, 2015/2016, indicating a negative bias of the wave models. This discrepancy between the two model vs. OSP comparisons was not solved in the poster, which had a suggestion for future work "Continue [$m_4$] evaluations using data provided by Dr. Eric Terrill (UCSD). Evaluate other months. Determine reason for inconsistent [$m_4$] bias of ST4/WW3: winter 2015/2016 vs. 2016/2017 (biofouling?)". This turned out to be the correct explanation: the OSP Datawell buoy was biofouled during the 2016/2017 period (Appendix B). In summary,

- SWAN/ST6 vs. UW/Datawell buoy, not biofouled, Nov.-Dec. 2015: bias=-11%
- WW3/ST4 vs. UW/Datawell buoy, not biofouled, Nov.-Dec. 2015: bias=-12%
- WW3/ST4 vs. UCSD/SIO/CORDC buoys, Nov. 24 2016 – Jan. 31 2017: bias=+24%
- WW3/ST4 vs. UW/Datawell buoy, biofouled, Nov. 24 2016 – Jan. 31 2017: bias=+28%

However, this only deepened the mystery: *why was the WW3 vs. UCSD/SIO/CORDC comparison consistent with the WW3 vs. biofouled OSP comparison, but not consistent with the WW3 vs. non-biofouled OSP comparison*? We did not investigate further, leaving this as a "cold case" until this year (2025).

## 3.2. Results: $m_4$, to 0.58 Hz, UCSD/SIO/CORDC buoys

The global WW3 hindcasts and comparisons to the UCSD/SIO/CORDC drifting buoys performed in 2017 were repeated using modern, consistent WW3 code and settings described in Section 2.3.1. In Figure 3, we compare $m_4$ , using $f_1$ and $f_2$ values consistent with those used in Rogers (2017): $f_1$=0.042 Hz and $f_2$=0.58 Hz. The left panel evaluates the hindcast using NAVGEM forcing, consistent with forcing used by Rogers (2017). The center and right panels use ERA5 forcing. The left and center panel show the results for the time period consistent with Rogers (2017), 24 November 2016 to 31 January 2017. The right panel shows the full time period for which we have the UCSD/SIO/CORDC observations, 24 November 2016 to 11 April 2017. This is observational dataset #2 described in Section 2.2 with locations shown in Figure 1.

Several error metrics are indicated in the plots: "n" is the number of co-locations, "rmse" is the RMS error, "CC" is the Pearson correlation coefficient, "SI" the scatter index, and "nbias" is the normalized bias. The scatter index is the standard deviation of the errors (i.e., de-biased RMSE) normalized by the mean of the observations.

[9] The poster was originally presented at the "Waves In Shallow Environments" meeting in Vancouver, Canada, 14-18 May 2017. The poster at the link provided was updated 27 June 2017, adding the comparison with the UCSD dataset and a discussion of physics parameterizations in the wave model "SWAN" (Simulating Waves Nearshore, Booij et al. 1999).

Recall that in Rogers (2017), this comparison showed a +24% positive bias in WW3. In all cases shown here, there is a similar, positive bias (+26%, +29%, and +29%). Thus, the primary conclusion to be drawn from Figure 3 is that the significant positive bias is not sensitive to model forcing, model version, or model settings.

Secondary conclusions, less relevant to the purpose of this report, can be made from Figure 3: the simulations using ERA5 show slightly better accuracy, e.g., scatter index is reduced from 0.22 to 0.21, and CC is increased from 0.90 to 0.91. The correlation of 0.90 matches the CC shown in Rogers (2017), while, disappointingly, the SI is slightly worse (0.22 vs. 0.20).

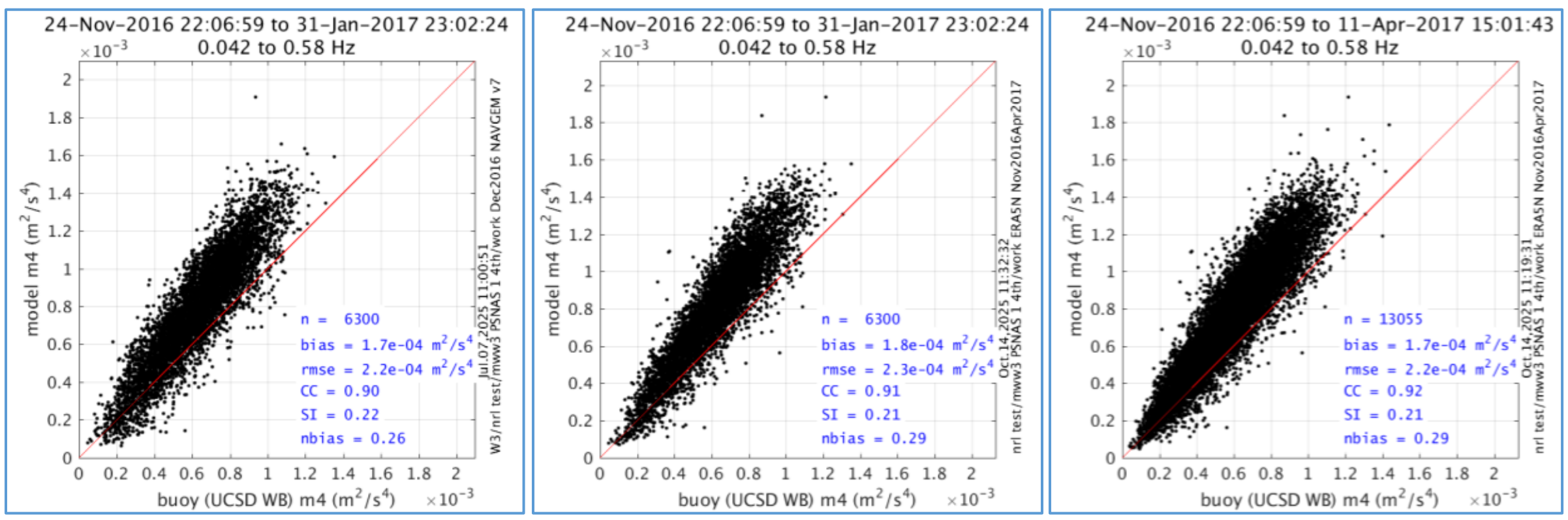

Figure 3. Scatter plots for $m_4$, WW3 vs. UCSD/SIO/CORDC buoys. Left panel: WW3 with NAVGEM forcing, 24 November 2016 to 31 January 2017. Center panel: WW3 with ERA5 forcing, 24 November 2016 to 31 January 2017. Right panel: WW3 with ERA5 forcing, 24 November 2016 to 11 April 2017.

### 3.3. Results: $H_{m0B}$, 0.2 Hz to 0.58 Hz, UCSD/SIO/CORDC and OSP buoys

In Figure 4, we repeat the comparison of the prior section, except this time, comparing $H_{m0B4}$ with $f_1$=0.2 and $f_2$=0.58 Hz. The consistent, positive bias is seen again, though it is smaller for this parameter (+14%, +15%, and +15%).

In Figure 5, we make a similar comparison of an ERA5-forced WW3 against the UW/APL buoy at Ocean Station Papa. This is observational dataset #1 described in Section 2.2 with location shown in Figure 1. A different time period is used, 16 May 2022 to 17 May 2023, since the UW/APL buoy was biofouled during the time period used in Figure 4. The number of co-locations is larger in Figure 5 (17.5k) than in Figure 4 (13k). Correlation is much better (0.97 vs. 0.88 and 0.89), though some of this improvement may be associated with the different location and time period. The bias is low (-4% instead of +15%), and the qualitative appearance of the scatter plot also indicates good skill.

Taken together, the results are consistent with those of Rogers (2017). That is to say, the evaluations against the UCSD/SIO/CORDC buoys indicate positive bias of WW3 at high frequencies, while evaluations against the OSP buoy indicate small bias and generally high accuracy of WW3 at high frequencies.

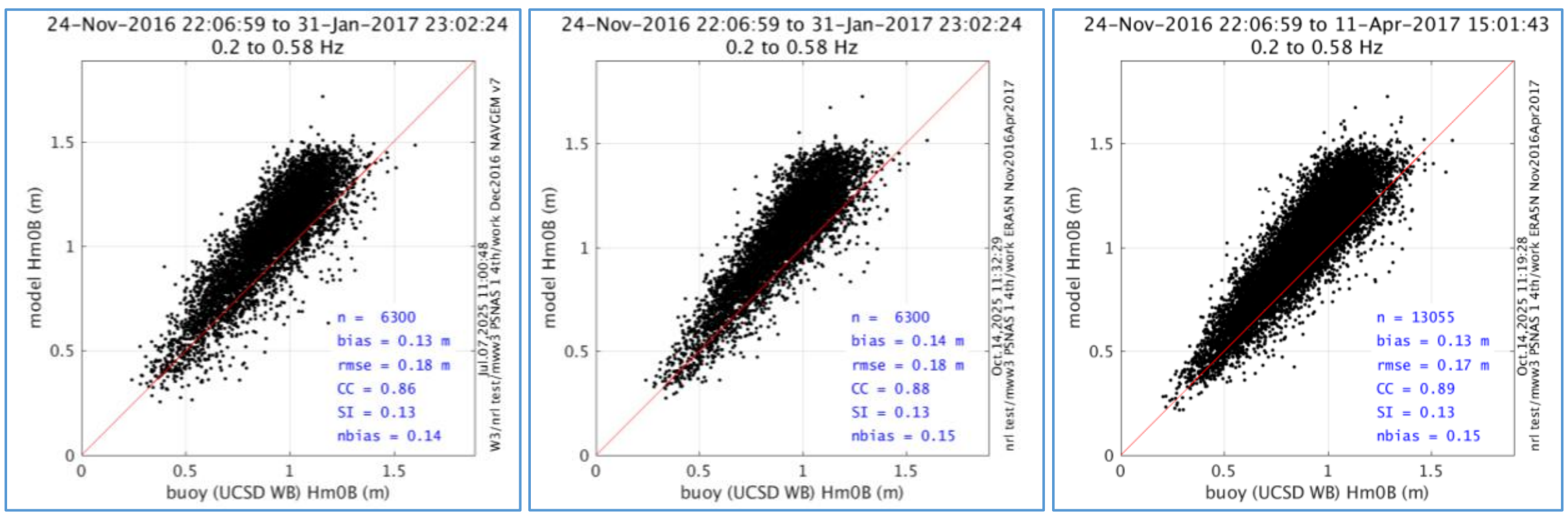

Figure 4. Like Figure 3, but showing scatter plots for band height $H_{m0B}$ computed from 0.2 to 0.58 Hz.

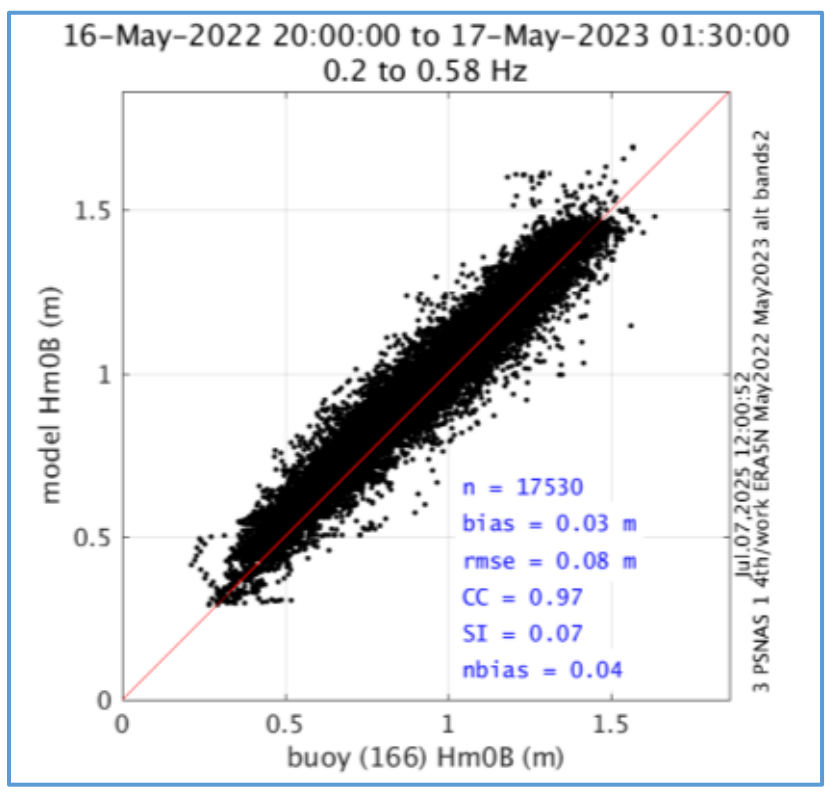

Figure 5. Like Figure 4, but showing WW3 vs. UW/APL buoy at OSP, 16 May 2022 to 17 May 2023.

### 3.4. Results: $H_{m0B}$, 0.2 Hz to 0.654 Hz, Sofar buoys

With the apparent discrepancy between the WW3 evaluations against the UCSD/SIO/CORDC buoys vs. WW3 evaluations against the OSP buoy, it is useful to introduce a third buoy dataset. In fact, an existing comparison was available, presented at the CNMOC-Sofar Technical Exchange Meeting July 17-18, 2023 (Rogers 2023). The most relevant part of that presentation is reproduced in Figure 6 and Appendix C here. [The full set of slides are available using this link.] Figure 6 compares the spatial distribution of the bias in $H_{m0B,4}$ of two WW3 hindcasts for February 2022, relative to $H_{m0B,4}$ computed using buoy data provided by Sofar. Thus, the observational dataset is a *one-month subset* of the nine-month dataset #3 described in Section 2.2 with locations shown in Figure 2. The first WW3 hindcast uses NAVGEM forcing and is equivalent to what was shown in Rogers (2023). The second uses ERA5 forcing, consistent with most other WW3 simulations presented in this report. Here, $f_1$=0.2 Hz and $f_2$=0.654 Hz. Thus $f_2$ is a bit higher than that used in the prior section[10]. Rogers (2023) showed that there was no systematic positive bias in bands 1, 2, and 3, but there is a systematic positive bias in band 4, reproduced in Figure 6. This outcome is the same for both simulations.

[10] The lowest $f_{max}$ for Sofar buoys is 0.654 Hz (82% of data), and in fact, some buoys report $E(f)$ to 1.25 Hz (18% of data). With the former dataset, the frequency grid is very irregular: with the last three bins being [0.469 0.498

In Figure 7, we present the overall results for the two simulations, using scatter plots for $H_{m0B,4}$. The overall bias is +6% using NAVGEM forcing and +9% using ERA5 forcing.

To summarize the $H_{m0B,4}$ results shown so far of the cases where WW3 is forced with ERA5:

- vs. UCSD/SIO/CORDC, $f_1$=0.2 Hz, $f_2$=0.58 Hz, WW3 has bias of +15%
- vs. OSP buoy, $f_1$=0.2 Hz, $f_2$=0.58 Hz, WW3 has bias of +4%
- vs. Sofar buoys, $f_1$=0.2 Hz, $f_2$=0.654 Hz, WW3 has bias of +9%

Thus, the conclusions based on the Sofar comparison regarding WW3 bias in high frequencies is intermediate between evaluation using the UCSD/SIO/CORDC buoys (large positive bias) and the evaluation using the OSP Datawell buoy (small positive bias).

---

0.654] Hz. With the latter dataset, the spacing is regular, $\Delta f = 0.0098\ Hz$, with the last three bins (rounded to three decimals) being [1.226 1.235 1.245] Hz.

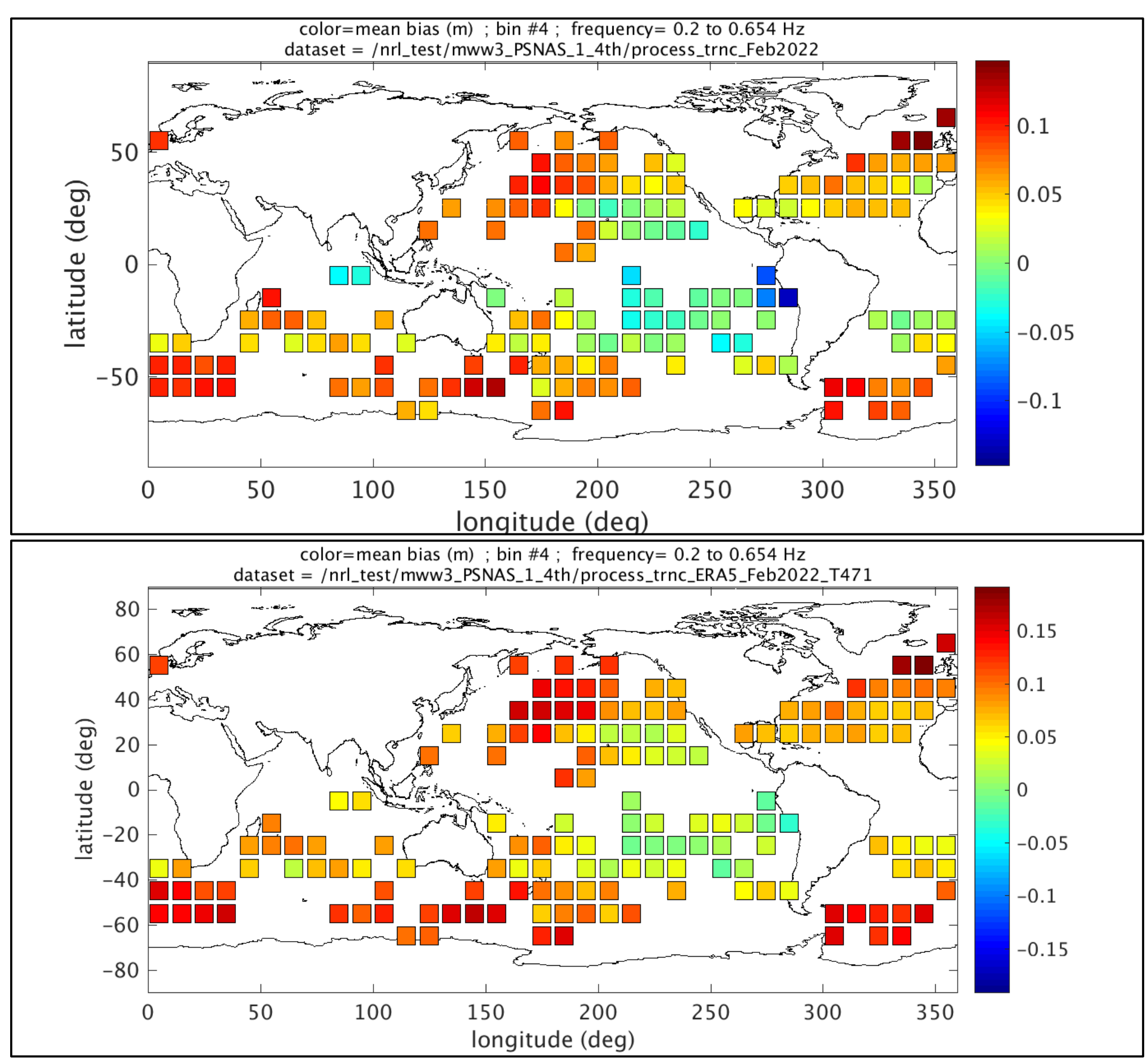

Figure 6. Bias (m) in $H_{m0B}$ of WW3 hindcast relative to Sofar buoys for the month of February 2022. $H_{m0B}$ is computed from 0.2 to 0.654 Hz. Top panel: WW3 with NAVGEM forcing. Lower panel: WW3 with ERA5 forcing. See also Figure 24.

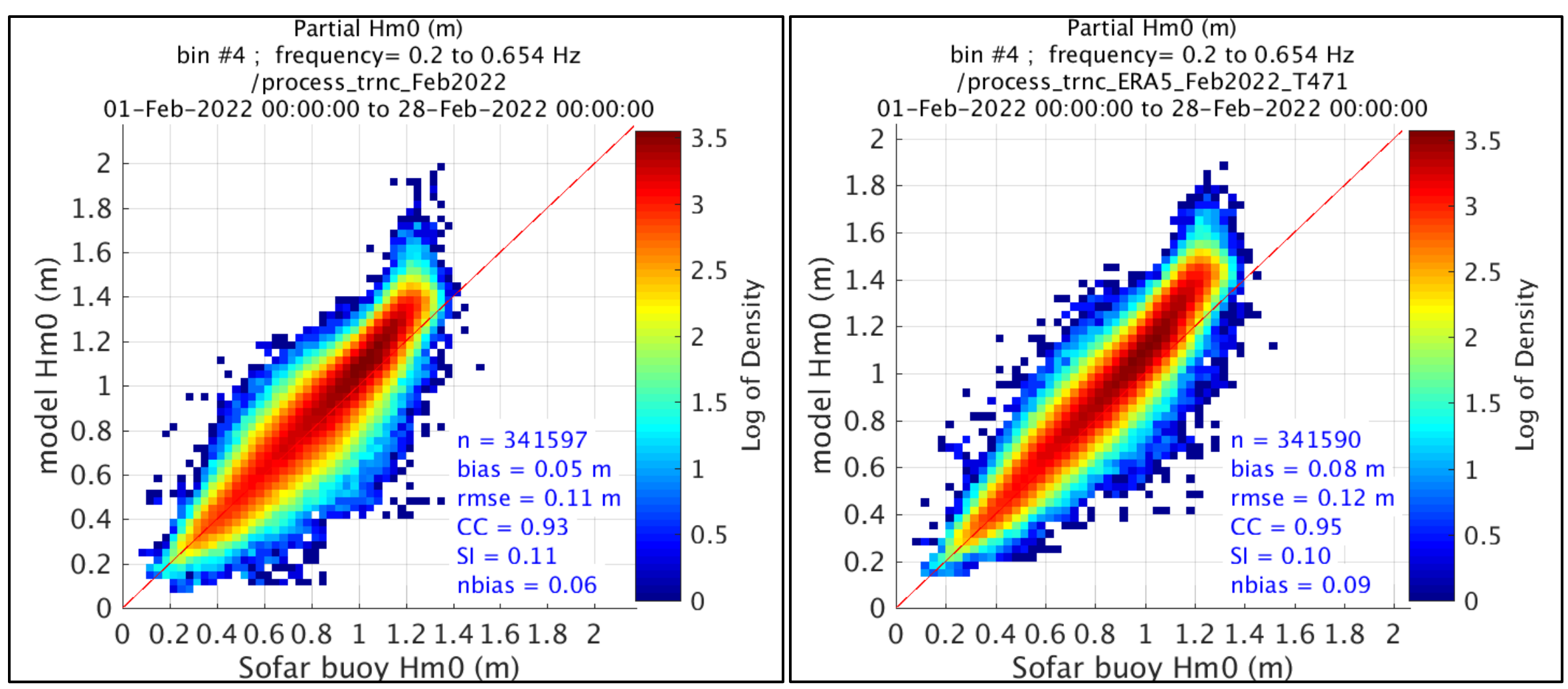

Figure 7. Scatter plot of $H_{m0B4}$ of WW3 hindcast relative to Sofar buoys for the month of February 2022. $H_{m0B4}$ is computed from 0.2 to 0.654 Hz. Left panel: WW3 with NAVGEM forcing. Right panel: WW3 with ERA5 forcing. See also Figure 25.

# 4. Wind speed-based evaluation

## 4.1. Results: $\boldsymbol{H_{m0B}}$, 0.29 Hz to 0.58 Hz

### *4.1.1.* $\boldsymbol{H_{m0B}}$ *vs.* $U_4$

In Figure 8, we compare two $H_{m0B}$ dataset against co-located wind estimates, for the period 24 November 2016 to 11 April 2017. Frequency limits are $f_1$ =0.29 Hz and $f_2$=0.58 Hz. The first $H_{m0B}$ dataset is computed from the UCSD/SIO/CORDC buoy data. The second is from co-located results from a WW3 simulation forced with ERA5. Wind estimates are in the form of $U_4$ approximated as $U_4 \approx 0.91U_{10}$, where $U_{10}$ is the 10-m neutral wind speed from ERA5. Scatter is larger at low wind speeds, particularly for the WW3 comparison, but the mean trend is robust for both datasets. Consistent with Figure 4, this comparison indicates that WW3 has a positive bias relative to the UCSD/SIO/CORDC buoys.

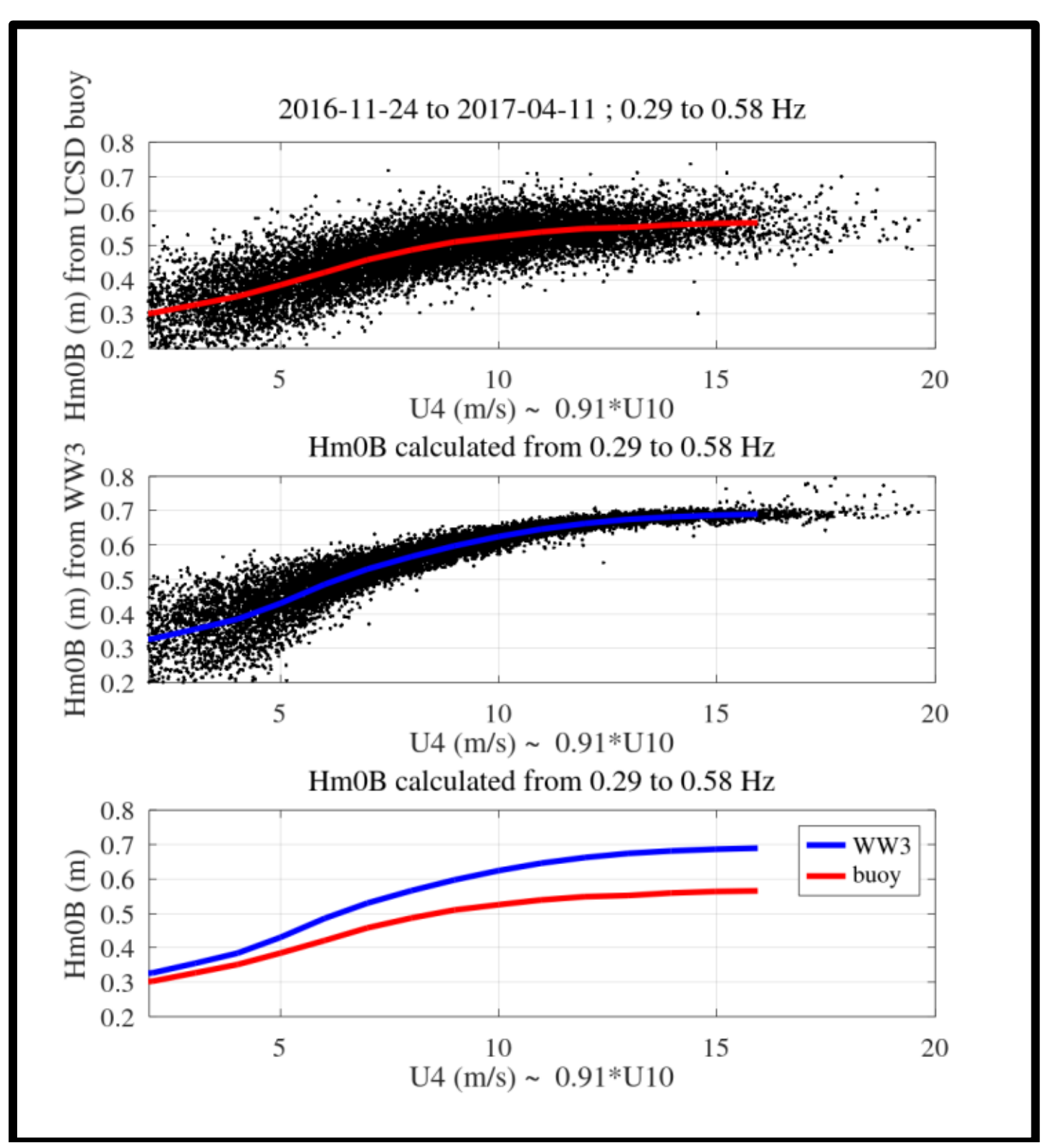

Figure 8. $H_{m0B}$ vs. $U_4$ , where the latter is approximated from $U_{10N}$. Black dots show match-ups. Colored lines show mean value of matchup at each wind speed bin. Bins with fewer than 100 values are not included in the mean-value lines. Top panel: UCSD/SIO/CORDC buoys. Center panel: WW3 at the locations of the UCSD/SIO/CORDC buoys. Lower panel: mean values of UCSD/SIO/CORDC buoys and WW3 on the same axis.

In Figure 9, we take the two mean trend lines from Figure 8, and overlay values measured at OSP. The OSP $H_{m0B}$ values are mean trend lines from the APL/UW buoy during months directly following buoy deployment, selected to avoid any risk of showing biofouled spectra. Bins with fewer than 30 values are not included in the OSP mean-value lines. The OSP $U_4$ values are taken from observations at a meteorological buoy maintained by NOAA/PMEL. The plot suggests that WW3 has a negligible bias relative to the OSP buoy.

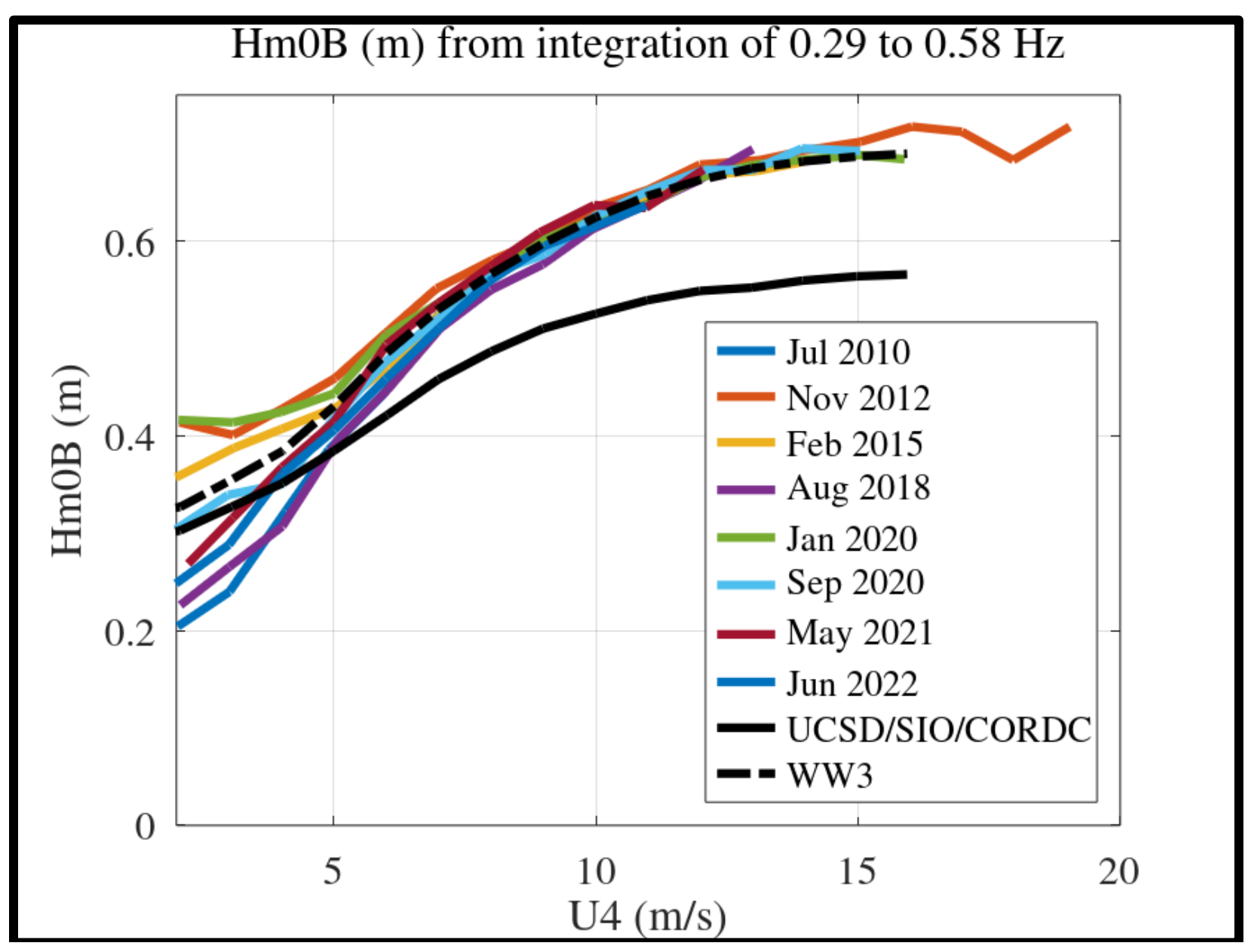

Figure 9. $H_{m0B}$ vs. $U_4$. Colored lines are taken from the first month of various deployments of the UW/APL OSP buoy, plotted against $U_4$ measured by an anemometer on the PMEL buoy. Black lines are the mean lines from Figure 8, where $U_4$ is approximated from $U_{10N}$. The solid black line corresponds to the UCSD/SIO/CORDC buoys.

#### *4.1.2.* $\boldsymbol{H_{m0B}}$ *vs.* $U_{10N}$

Figure 10 compares six $H_{m0B}$ datasets. The left column of plots show, from top to bottom:

1) the UW/APL buoy at OSP May 2022 to May 2023,
2) the UCSD/SIO/CORDC buoys period 24 November 2016 to 11 April 2017,
3) the Sofar buoys during February 2022. A quality control has been applied: observations for which $H_{m0B} < 3$ cm and $U_{10N} > 2.5$ m/s are considered spurious and omitted. This is discussed further below.

The right column of plots show, from top to bottom:

4) WW3 co-located with (1).
5) WW3 co-located with (2).
6) WW3 co-located with (3).

Frequency limits are the same as those used in the prior plots (Section 4.1.1). For all cases, wind speed on the horizontal axis is $U_{10N}$ from ERA5, and the WW3 forcing is also from ERA5. Bins with fewer than 100 values are not included in the mean-value lines.

Figure 11 shows a seventh $H_{m0B}$ dataset, based on Sofar observations September 2021 to May 2022. This is observational dataset #3 described in Section 2.2 with locations shown in Figure 2. Bins with fewer than 100 values are not included in the mean-value line. We did not run a WW3 simulation for co-location with this entire dataset. The mean trend line shown here is used later (Figure 13).

In Figure 11, we there are an unusually high number of co-locations falling along the horizontal axis, indicating that some of the Spotter buoys are reporting $H_{m0B}$ values that are much smaller than expected. Figure 12 shows where these measurements are made, selecting cases with $H_{m0B} < 3$ cm and $U_{10N} > 2.5$ m/s. Many of the locations are near coastlines, suggesting that the buoys are grounded. Other locations are in the Antarctic, suggesting that the spectra measured by the buoys are damped by sea ice. Using the criterion of Figure 12, only 4715 of the observations are affected, which is 0.14% of the observations. Thus, it has a barely perceptible impact on the mean trend line[11].

Figure 13 clearly shows that $H_{m0B}$ from the WW3 models are consistent with those from the moored OSP buoy but are higher than the values from drifting buoys. The drifting buoy values (UCSD/SIO/CORDC and Sofar) are consistent with each other. The similarity of the three WW3 datasets indicates that according to WW3, the impact of discrepancy in time/location is negligible for these mean-trend evaluations.

Though it is not a primary concern of this report, the non-monotonic trend in Figure 11 is interesting, with $H_{m0B4}$ peaking around $U_{10N}$=15 m/s. The non-monotonic behavior is not a surprise since wind stress is known to saturate and even decline at high wind speeds. However, this normally is seen at larger wind speeds, e.g., with Jarosz et al. (2007), the saturation of wind stress occurs at $U_{10}$>40 m/s and with Hwang (2011), it occurs at $U_{10}$>50 m/s. Davis et al. (2023) show something similar for mean square slope, with a limit occurring at $U_{10}$>30 m/s. This suggests that the wind speed of saturation strongly depends on the parameter that is being evaluated (wind stress, $H_{m0B4}$, MSS). The model does not reproduce the non-monotonic behavior; this is especially evident in the Sofar comparisons in Figure 10.

[11] Though the impact is barely perceptible, it is worthwhile to indicate exactly what we are showing: In Figure 11, the mean trend line does not include the QC. In Figure 10 (lower left panel), the mean trend line includes the QC. In Figure 13, both mean trend lines for the Sofar data includes the QC.

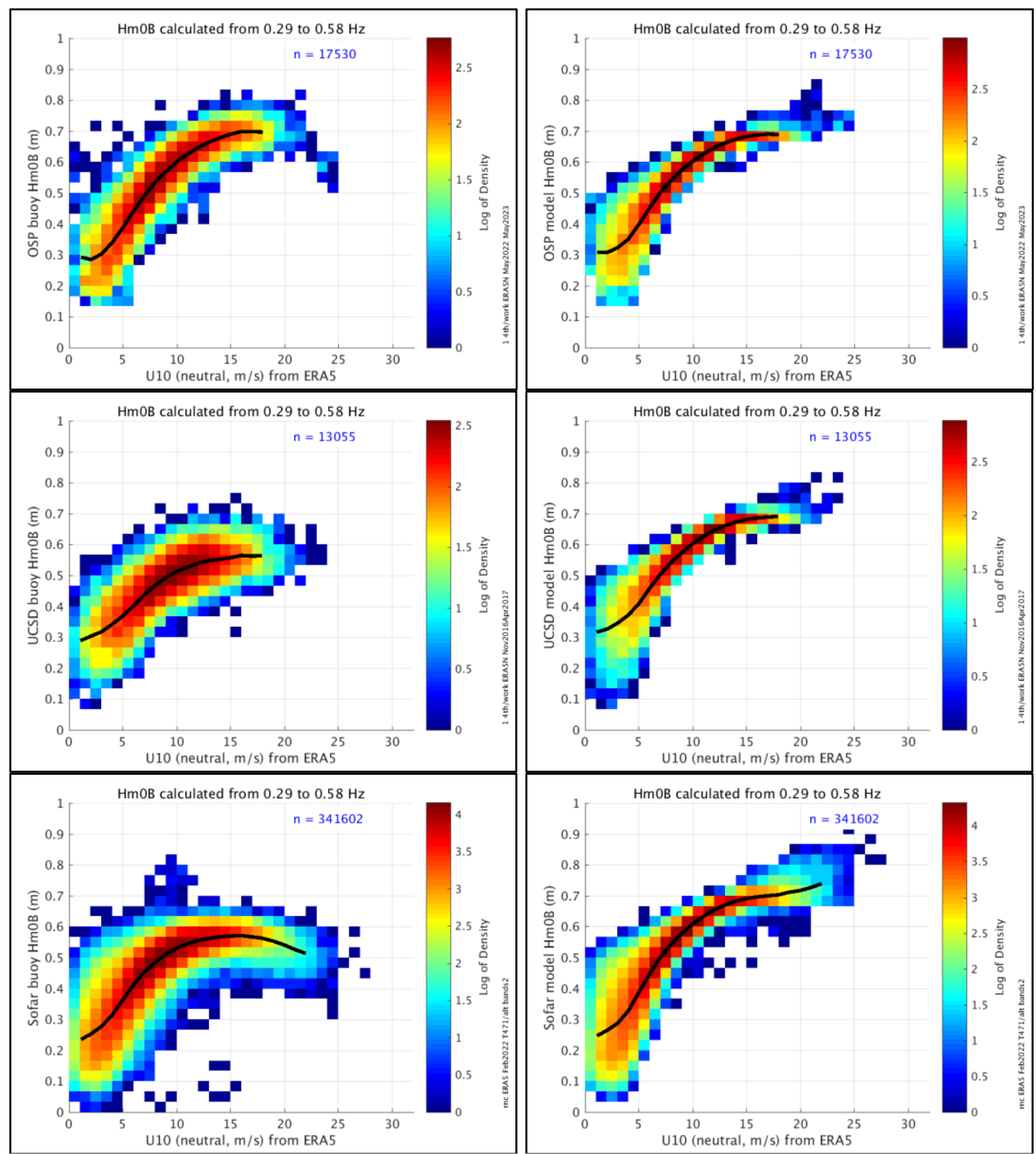

Figure 10. $H_{m0B}$ vs. $U_{10N}$. $H_{m0B}$ is computed from 0.29 to 0.58 Hz. The color bar indicates the density of scatter plot match-ups. Black lines are the mean values. Top panels: OSP location. Center panels: UCSD/SIO/CORDC drifting buoy locations. Bottom panels: Sofar drifting buoy locations (February 2022). Left panels: $H_{m0B}$ is taken from buoy observations. Right panels: $H_{m0B}$ is taken from WW3 at the buoy locations.

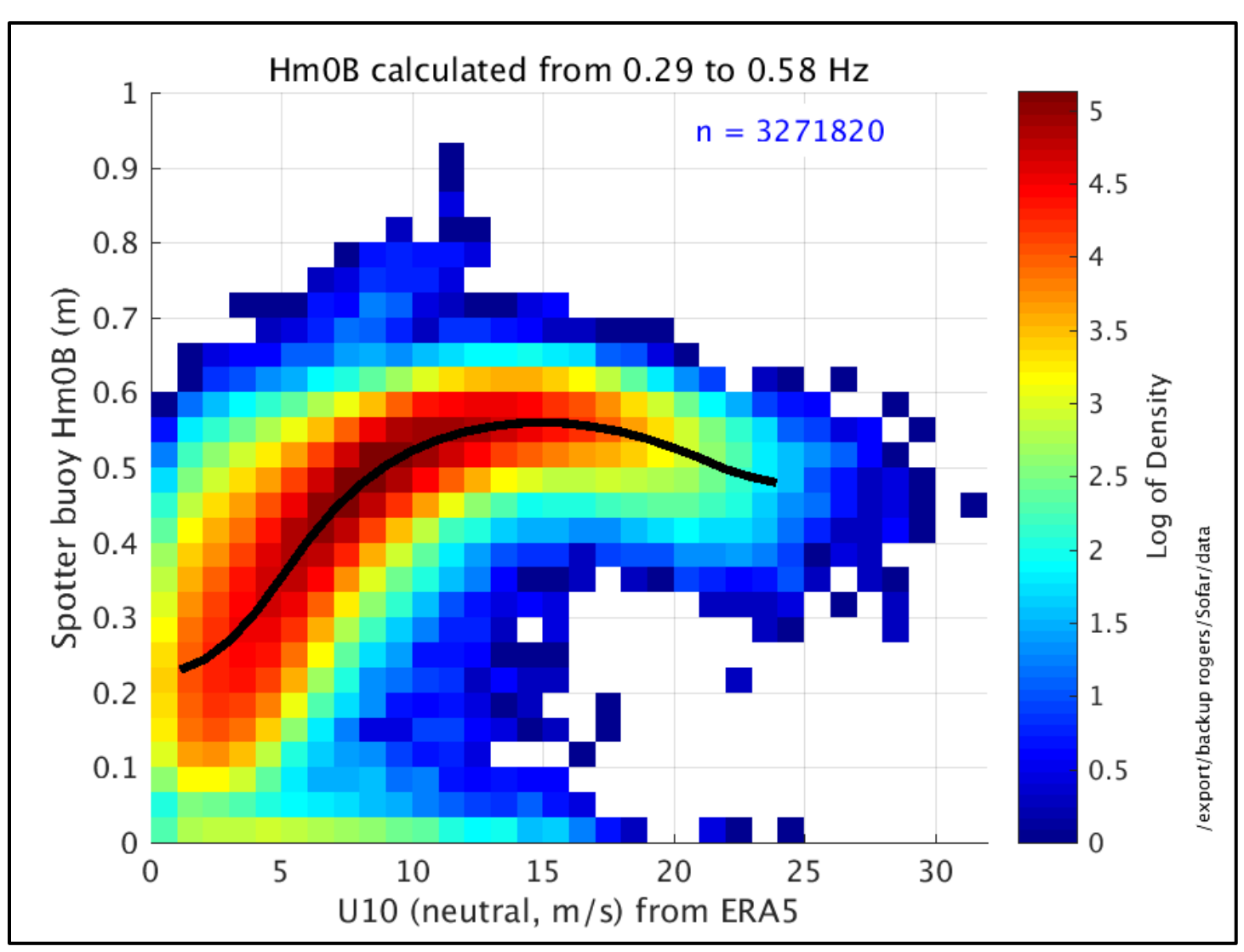

Figure 11. Like Figure 10, but $H_{m0B}$ is taken from Sofar buoy observations, nine-month dataset.

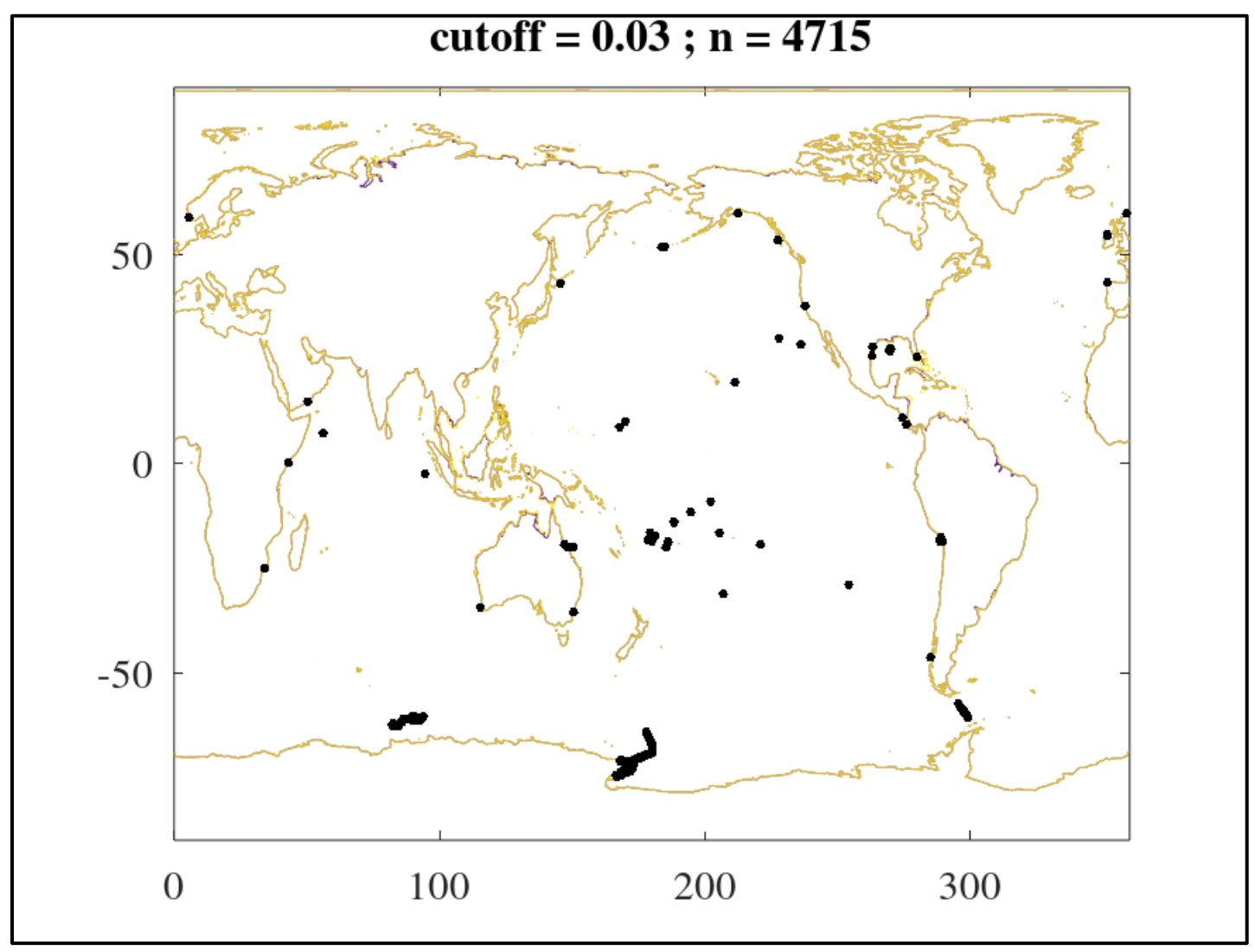

Figure 12. Geographic distribution of Sofar co-locations in Figure 11 with $H_{m0B} < 3$ cm and $U_{10N} > 2.5$ m/s.

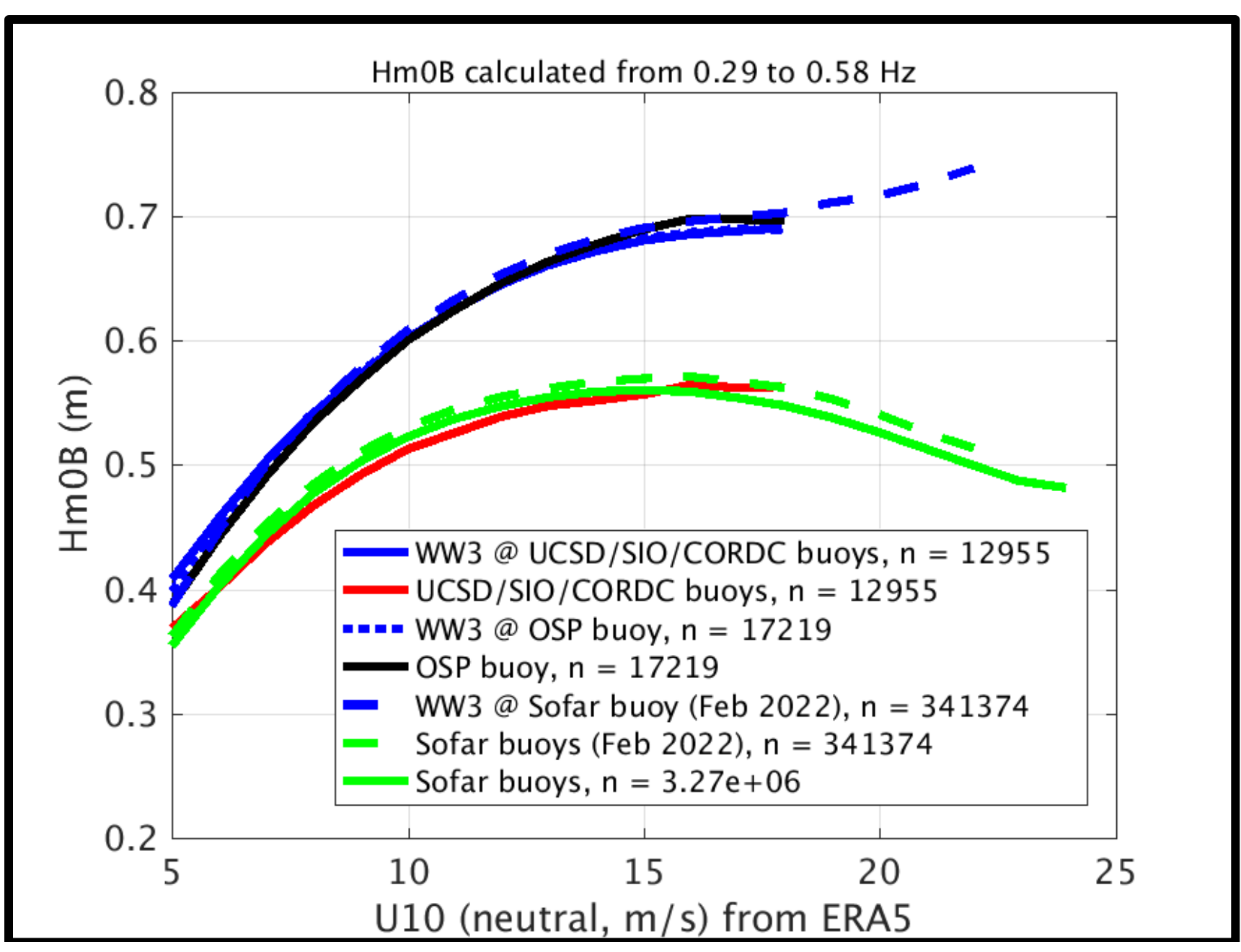

Figure 13. Mean-value lines taken from the seven density scatter plots of Figure 10 and Figure 11. The red line corresponds to the UCSD/SIO/CORDC buoys.

### 4.2. Results: *E(f)*, 0.4 Hz

In Figure 14, we repeat the comparison of the prior section, except that we are evaluating based on the spectral density at 0.4 Hz, in place of $H_{m0B4}$. Since the February 2022 WW3-Sofar match-ups did not contribute meaningfully to Figure 13, they are omitted here. The Sofar data are quality-controlled by removing cases where spectral density $E(f)$ is less than $1 \times 10^{-4}$ m2/Hz and wind speed $U_{10N} > 2.5$ m/s. As with the QC applied to the Sofar $H_{m0B4}$, this reduces the population by 0.14% (from $3272 \times 10^{-3}$ to $3267 \times 10^{-3}$) and has a barely perceptible effect on the mean trend line.

In Figure 14, we introduce a parametric spectrum as a reference line. This is described in Appendix D. It is based on work by Toba (1973), Forristall (1981), Battjes et al. (1987), Toba (1988), and Holthuijsen (2007).

The outcome of Figure 14 is largely consistent with Figure 13.

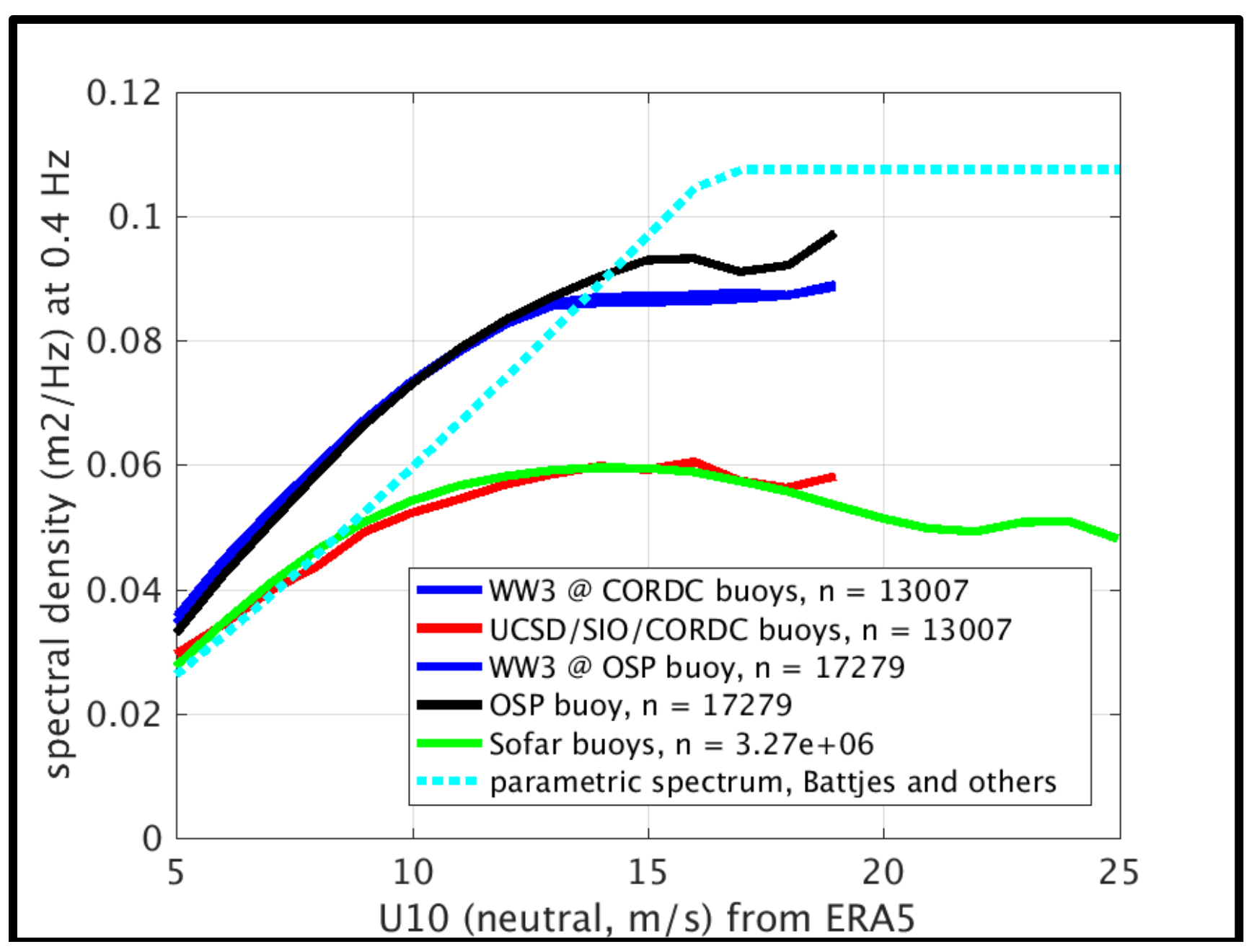

Figure 14. Like Figure 13, except that the vertical axis is $E(f)$ for $f$ =0.4 Hz. The Phillips spectrum level is described in the text. The OSP (CDIP/Datawell) buoy dataset is May 2022 to May 2023.

It is reasonable to wonder if there is something peculiar about the OSP buoy, relative to its siblings at other locations. To answer this, we performed similar co-locations of $E(f)$ from CDIP Datawell buoys with ERA5 $U_{10N}$ values; see Figure 15. Here, the CDIP time series are at a 30-minute interval, "up to" 32 months duration, May 2022 to December 2024, most are shorter than 32 months, due to outages: the numbers of colocations are indicated for each buoy in the legend. This is observational dataset #4 described in Section 2.2 with locations shown in Figure 2. Note that the OSP time series here is longer ($n$ =29k) than was used in Figure 14 ($n$ =17k), and so the mean trend lines are not expected to match exactly. Bins with fewer than 45 values are not included in the mean-value lines.

The comparison indicates that the OSP buoy is <u>not</u> an outlier relative to the other CDIP buoys.

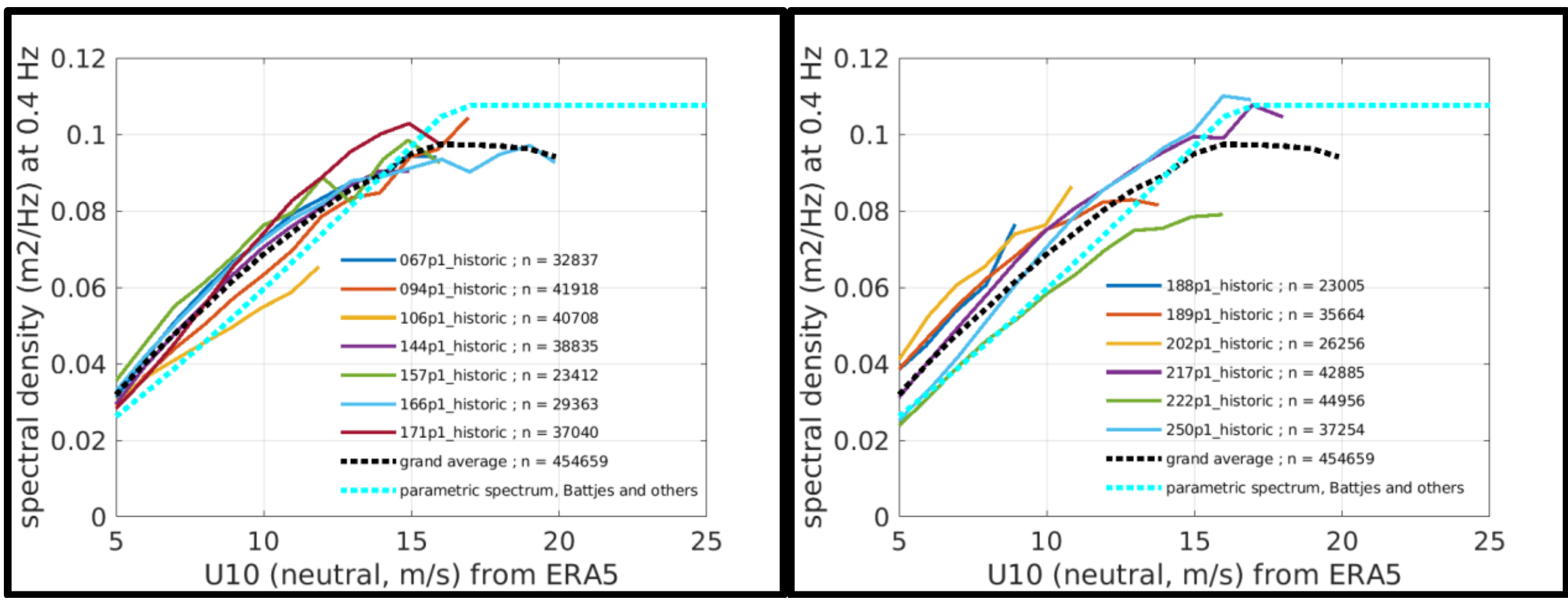

Figure 15. Like Figure 14, except showing results for other, individual CDIP buoys, May 2022 to December 2024. The buoy numbers and number of matchups are indicated in the legend. The left and right panel show two different sets of CDIP buoys. The organization (left vs. right) and order of the buoys is arbitrary.

By basing our analysis on $E(f)$ instead of $H_{m0B4}$, we make it possible to include buoys which do not report to the $f_2$=0.58 Hz used in our $H_{m0B4}$ calculations: namely, buoys of NOAA's National Data Buoy Center (NDBC). Most contemporary NDBC buoys report to 0.485 Hz. This comparison is made in Figure 16, using eight NDBC buoys, which are observational dataset #5 described in Section 2.2 with locations shown in Figure 2. Bins with fewer than 45 values are not included in the mean-value lines. The buoys are briefly summarized here:

- 41001 - deep, 3-m foam, SCOOP payload, 150 NM E of NC
- 41002 - deep, 3-m foam, SCOOP payload, 225 NM S of NC
- 42001 - deep, 3-m foam, SCOOP payload, "MID GULF" 180 NM S of LA
- 42002 - deep, 3-m foam, SCOOP payload, "WEST GULF" 207 NM E of TX
- 44014 - 49 m depth, 2.1-m foam, SCOOP payload, 64 NM E of VA
- 46001 - deep, 3-m foam w/seal cage, SCOOP payload, western Gulf of Alaska, 175 NM SE of Kodiak AK
- 46066 - deep, 3-m foam, SCOOP payload, 310 NM SSW of Kodiak, AK
- 51000 - deep, 3-m foam, SCOOP payload, 245 NM NE of Honolulu, HI

Above descriptions were retrieved from https://www.ndbc.noaa.gov/ October 29, 2025. We did not confirm that the hull/payload were the same during the data period, January to December 2024.

This comparison shows that the tail level of the NDBC buoys is significantly lower than that of the CDIP/DWR buoys. For example, $E(f)$ is 0.05 m$^2$/Hz near $U_{10N}$ = 12 m/s for six of the eight NDBC buoys, while most CDIP buoys have $E(f)$ near 0.08 m$^2$/Hz. In other words, the moored NDBC buoys appear to be more consistent with the drifting buoys (Sofar, UCSD/SIO/CORDC) than with the moored CDIP buoys. Of course, the reason for the discrepancy between CDIP buoys and the other buoys has not been established. It is likely that each (Sofar, CORDC, and NDBC) has a different reason for having a lower tail level than the CDIP buoys. Nevertheless, the simplest explanation is that the CDIP/DWR tail levels are wrong (artificially high) while the other datasets (Sofar, CORDC, and NDBC) have tail levels that are relatively accurate. An alternative, compound explanation is 1) that drifting buoys have lower tail levels because of the Doppler shift drifting as the buoys move downwind in the same direction as the waves and 2) moored NDBC buoys have a lower tail levels because their response is damped by their large size.

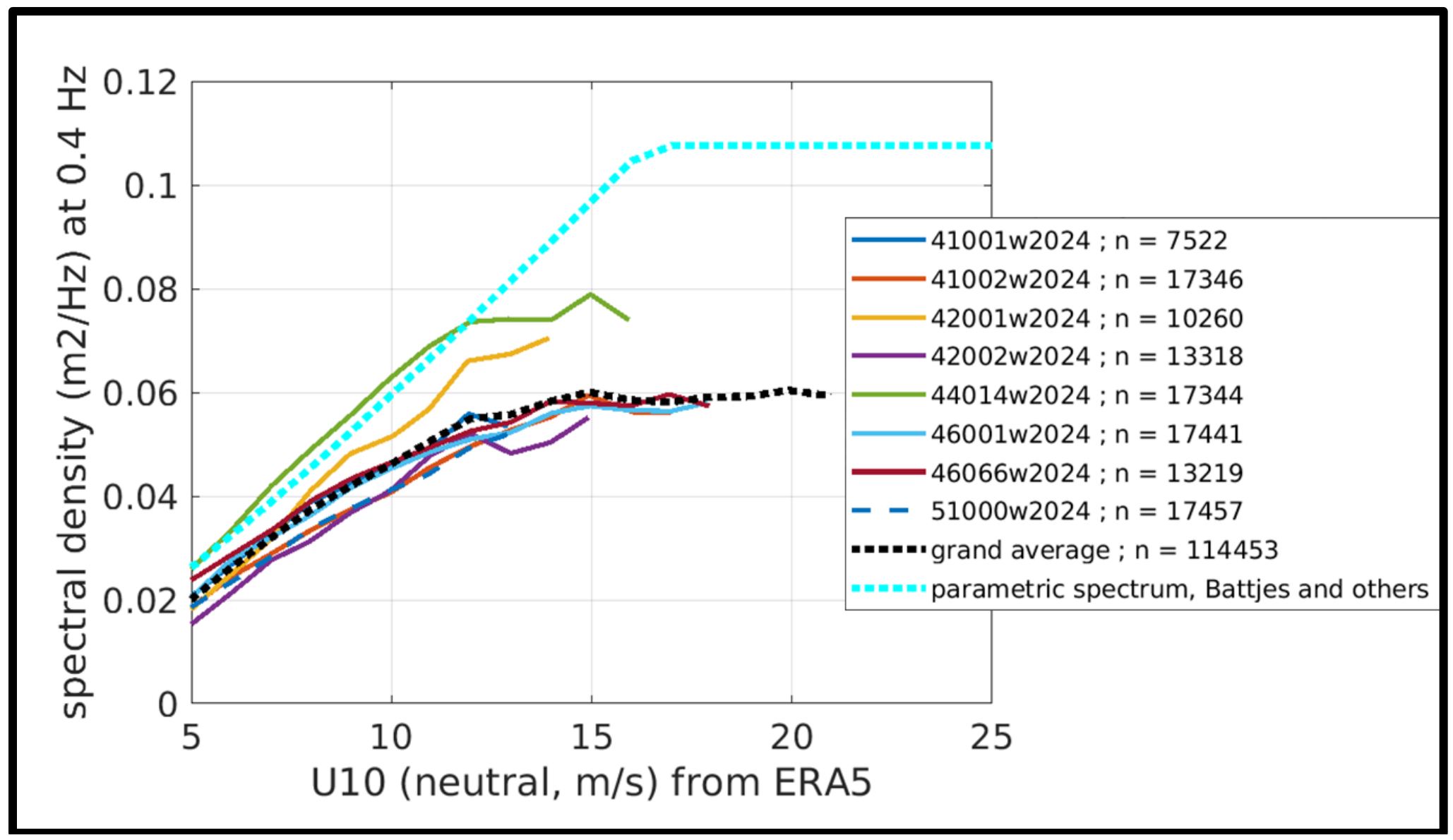

Figure 16. Like Figure 15, except showing results for NDBC buoys, January to December 2024.

In the case of NDBC buoy 44014 and CDIP buoy 171 (which has NDBC designation 44088), the two buoys are co-located, east of Virginia. Their tail levels are compared in Figure 17. Here, we use identical time series, 1 January 2024 to 13 June 2024 ($n$=3948 for both datasets). Unfortunately, NDBC buoy 44014 is somewhat of an outlier, having higher tail level than its siblings (see Figure 16), i.e., it is the NDBC buoy most similar to the CDIP buoys. It is also the only buoy among the eight NDBC buoys that is not moored in deep water; its depth is 49 m. Nevertheless, Figure 17 still shows higher tail level for the CDIP buoy than for the NDBC buoy, e.g., at $U_{10N}$=12 m/s, the former has $E(f)$~0.086 m$^2$/Hz and the latter has $E(f)$~0.073 m$^2$/Hz.

In Figure 18, we present the mean-trend lines of the observational datasets from prior plots.

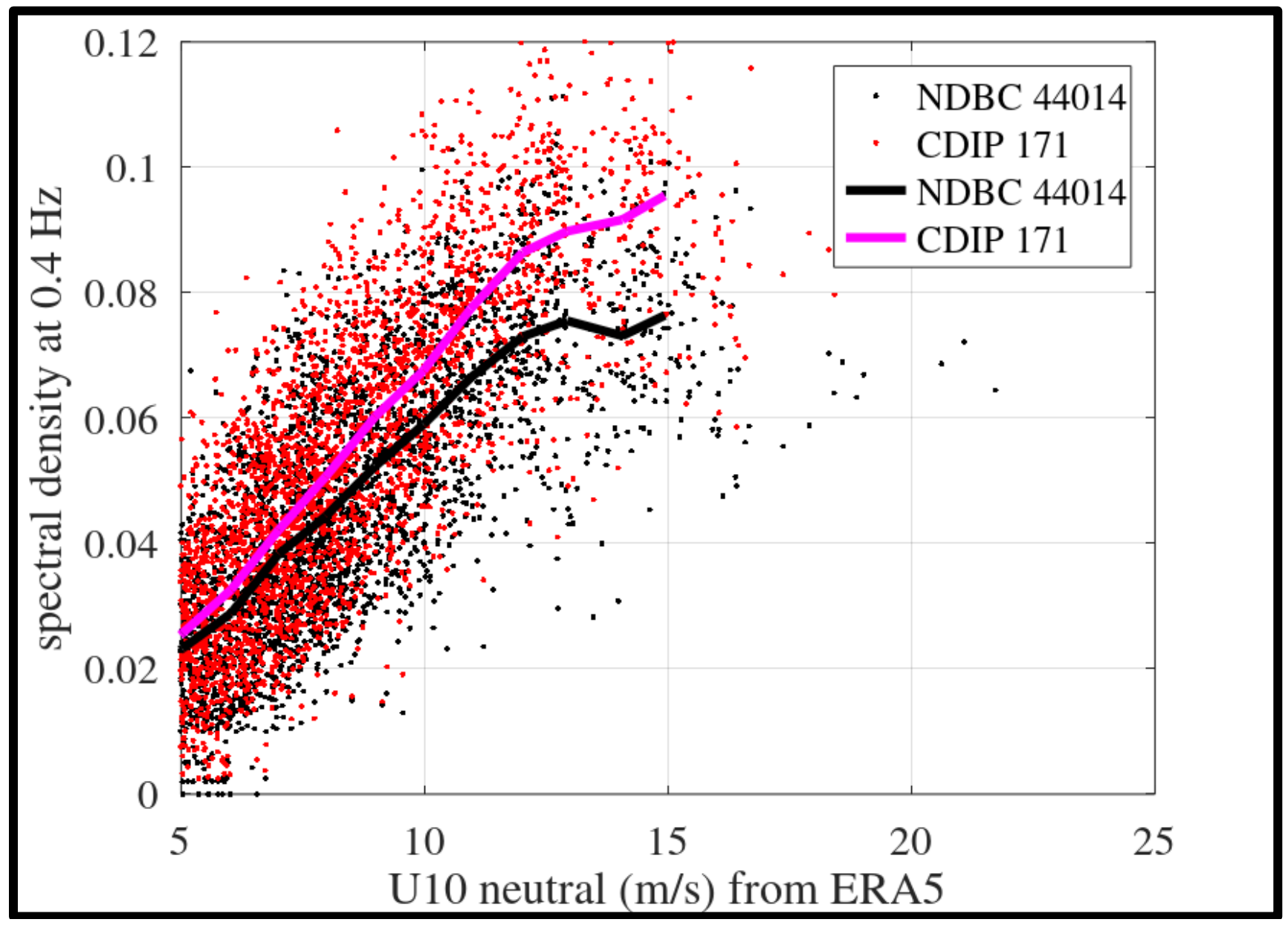

Figure 17. $E(f)$ for $f$ =0.4 Hz vs. $U_{10N}$. Direct comparison of NDBC 44014 and CDIP 171 against ERA5 wind speeds, 1 January 2024 to 13 June 2024. The dates of co-locations are identical for the two datasets, with 3948 co-locations each.

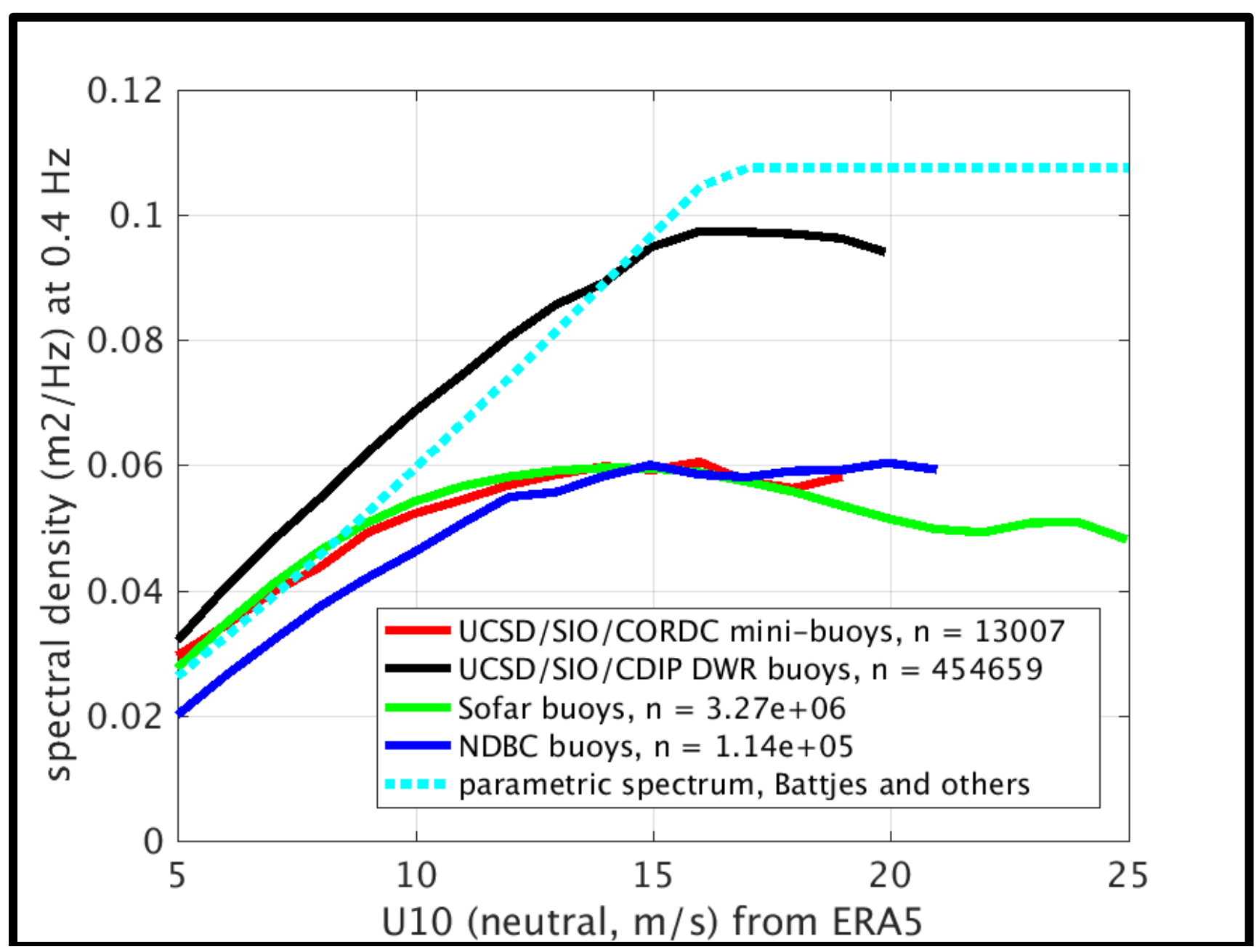

Figure 18. Mean trend lines of observational datasets presented in prior figures. The red line corresponds to the UCSD/SIO/CORDC mini-buoys, and the black line corresponds to the UCSD/SIO/CDIP DWR buoys.

## 5. High frequency tail slope from Rogers (2017) poster presentation

So far, we have evaluated tail level using the metrics of mean square slope, $H_{m0B4}$, and $E$(0.4 Hz). These are quantitative comparisons. It is also possible to evaluate the tail level by inspecting the tail slope, $\partial E(f)/\partial f$. Such comparisons tend to be more qualitative and/or subjective, but can provide useful insight. Here, we present tail level comparisons from the poster, Rogers (2017). Figure 19 compares the high frequency spectra from two buoys. The first buoy is the APL/UW/CDIP DWR at OSP. The second buoy is a 3-meter NDBC discus buoy south of Alaska. The most obvious takeaway from the comparison, remarked upon by Rogers (2017), is that there is reduction of the NDBC wave spectrum (i.e., increase in slope) between 0.4 Hz and 0.48 Hz which appears to be artificial, especially if we consider the DWR buoy as an exemplar for correct behavior. In Figure 20, the top panels illustrate the process of binning the DWR spectra by wind speed. This provides better clarity, as the lines (spectra) are not obscuring each other. The lower panels show the results from two models used by the USN, SWAN and WW3. If we consider the DWR spectra to be correct, then we are satisfied that both models reproduce this behavior: this was the sentiment of Rogers (2017).

In the DWR spectra, with the benefit of hindsight, we can see an indication of a problem: for the lower wind speeds, starting around 0.3 to 0.32 Hz, the spectral slope is artificially low, lower than the $f^{-4}$ slope. The defect is less obvious at the higher wind speeds, but we can infer a problem there also: since the spectrum is expected to transition from $f^{-4}$ to $f^{-5}$ at some transition frequency $f/f_p \approx 3$ (Tsagareli 2009)[12], and since the peak frequency $f_p$ is expected to be lower at high wind speeds, we might expect that slope in the 0.3 to 0.58 Hz range should be closer to $f^{-5}$ than $f^{-4}$ at higher wind speeds. The models provide this behavior, but the DWR buoy does not (Figure 20).

---

[12] If the reader does not have access to this reference, we refer the reader to Forristall (1981) and Kahma and Calkoen (1992), who argue for a transition from $f^{-4}$ to $f^{-5}$ when the wave age becomes small, with transition frequency $f_T \sim 0.29/U_*$ or $f_T \sim 7.8/U_{10}$ respectively. (These equations are equivalent if $U_{10} \sim 27U_*$)

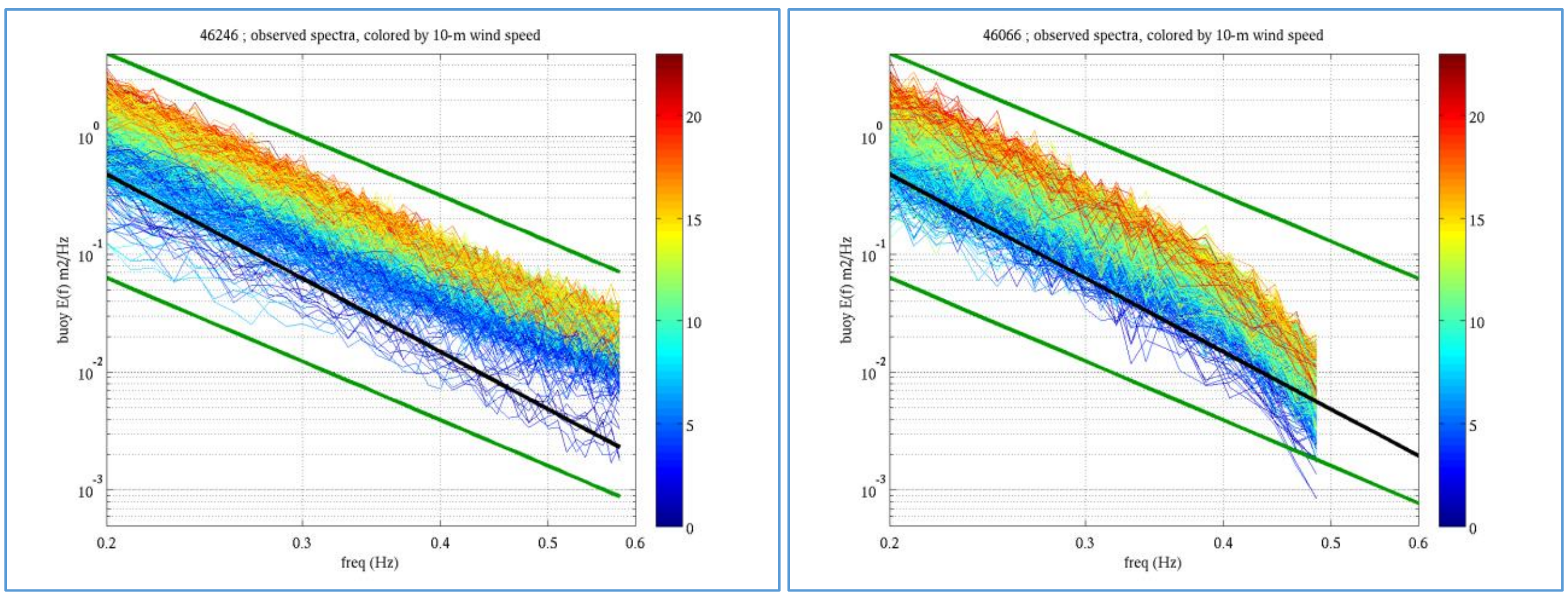

Figure 19. Plots taken from Rogers (2017): High-frequency spectra $E(f)$, where colors indicate the wind speed. The time period is 04 November 2015 to 25 December 2015. The thick green lines indicate $f^{-4}$ slope. The thick black lines indicate $f^{-5}$ slope. Left panel: results from OSP APL/UW/CDIP DWR buoy. Right panel: results from NDBC buoy 46066, a 3-m discus buoy, 310 nautical miles south-southwest of Kodiak, AK.

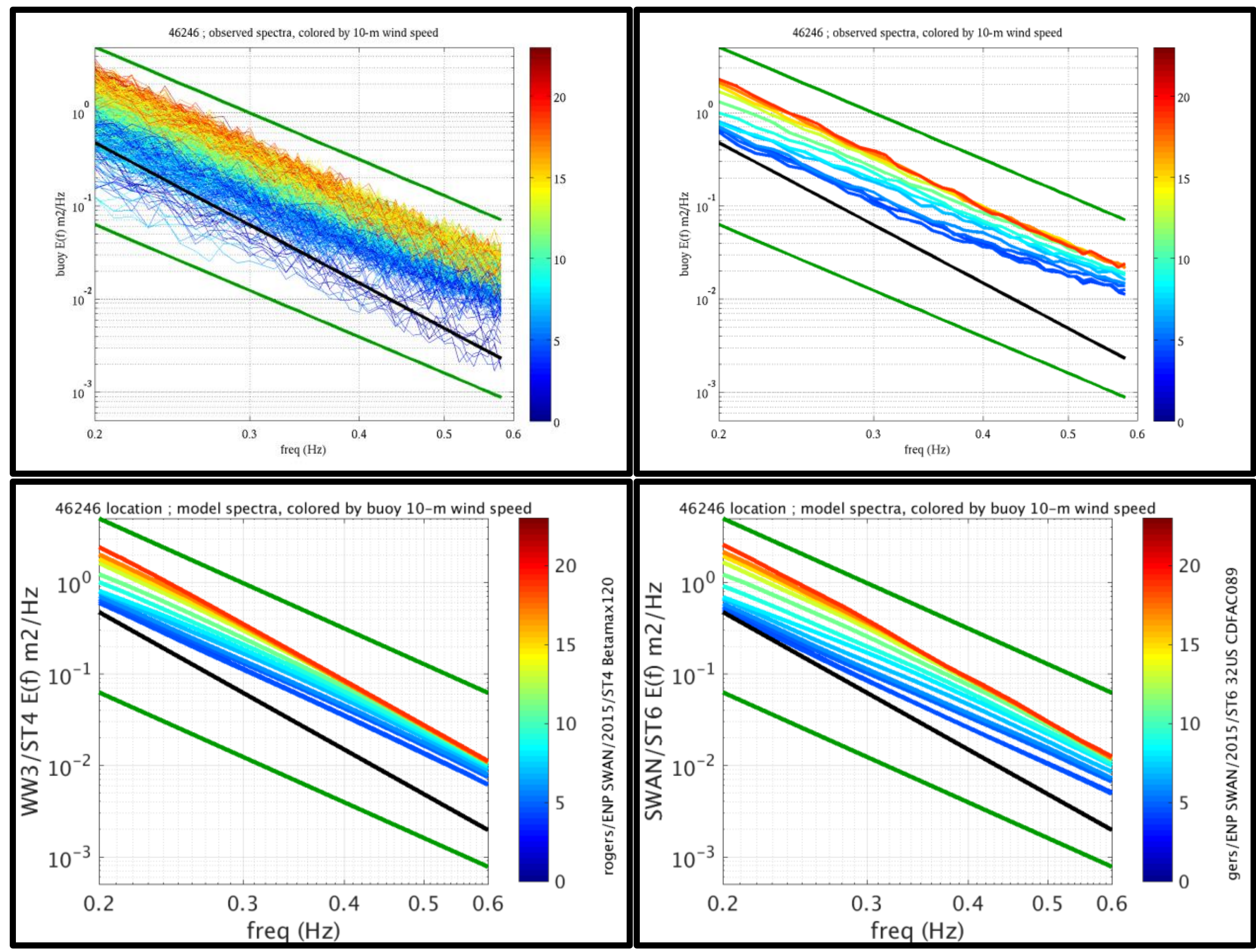

Figure 20. Similar to Figure 19, except as follows. All results correspond to the location of the OSP APL/UW/CDIP DWR buoy. Top left is identical to the left panel of Figure 19. Top right is the same data, with binning by wind speed. Lower panels are from models with binning by wind speed. Lower left: result from WW3 with ST4 physics parameterizations. Lower right: result from SWAN with ST6 physics parameterizations.

## 6. Discussion of possible explanations for the discrepancy

### 6.1. Doppler shift

Prior to the addition of the NDBC buoys to our comparison, there was a clear separation between our buoy datasets: the moored CDIP buoys report higher tail level than the drifting UCSD/SIO/CORDC and Sofar buoys. This led us to speculate that the discrepancy may be caused by Doppler shift. This would be a reassuring explanation, since it allows that both buoys are accurate, but just reporting from different frame of reference. With the addition of the NDBC buoys, Doppler shift can still be an explanation for CDIP vs. CORDC/Sofar, but we would require a separate explanation for the discrepancy between CDIP and NDBC. Moreover, it is possible that Doppler shift contributes to the discrepancy as described here, but is not the only cause. This will be explored in a follow-up study. In any case, a discussion of the Doppler shift follows.

Doppler shift is described by the Doppler equation, $\omega = \sigma + \vec{k} \cdot \vec{U}$. Here, $\omega$ is the "absolute frequency" measured from a fixed frame of reference, e.g., a moored buoy or a stationary pressure gage, $\sigma$ is known as the "relative" or "intrinsic" frequency, being the frequency from the frame of reference moving with the currents. The wavenumber vector is $\vec{k}$ and the current vector is $\vec{U}$. The wave model is formulated in terms of $\sigma$ and linear dispersion relates $\sigma$ to the scalar wavenumber $k$ as $\sigma^2 = gk\tanh(kh)$, where $h$ is water depth. The wave model does not account for vertical shear in the currents, meaning that the ocean is treated as a moving slab. In practice, model-derived Eulerian <u>surface</u> currents are typically used as input to a wave model, if they are accounted for at all. In reality, the current $\vec{U}$ is vertically sheared. Moreover, there is emerging clarity that $\vec{U}$ is not simply the Eulerian currents (as we commonly treat it), but is rather the Lagrangian mean current[13], which is the combined flow from the Eulerian currents and Stokes drift (see Pizzo et al. 2023 and Hara 2025).

For purposes of comparing to the wave model's $E(\sigma)$, it would be convenient if drifting buoys drifted exactly with $\vec{U}$, but unfortunately, due to windage, and drogue depth (and possibly other effects that we aren't thinking of), the buoys provide spectra in a third frame of reference, neither $E(\sigma)$ nor $E(\omega)$ (Battjes et al. 1987). Davis et al. (2025) denote this as $E(f_{ob})$. Collins et al. (2017) provide a comprehensive review.

Doppler shift does not affect total energy (or by association, significant wave height): $\int_0^\infty E(\sigma) d\sigma = \int_0^\infty E(\omega) d\omega$ . However, it is conceivable that Doppler shift could affect the energy in a particular band, because the reference frame will change the frequency at which energy appears: $\int_{2\pi f_1}^{2\pi f_2} E(\sigma) d\sigma \neq \int_{2\pi f_1}^{2\pi f_2} E(\omega) d\omega$
Here, $f_1$ and $f_2$ are the upper and lower limits of a band, e.g., $f_1$=0.29 Hz and $f_2$=0.58 Hz and frequencies are the frequency grid on which the buoy data are provided. The effect would be most important at higher frequencies, because those components have the largest values of $\vec{k} \cdot \vec{U}$.

As a thought experiment, take a case where buoy drift direction is the same as the windsea direction. The spectrum of the drifting buoy will down-shift frequency space, i.e., Doppler shift makes the wave period appear longer than that which would be measured from a fixed point. Since energy level drops with frequency (roughly at $f^{-4}$), this redshift reduces the energy in the (high frequency) $H_{m0B4}$ band. If, in a global/climatological sense, there is a significant positive correlation between buoy drift direction and the windsea direction (via mutual correlation with wind speed[14]), this thought experiment could work as an explanation for why drifting buoys report lower $H_{m0B4}$ than the moored buoys.

[13] This is also called the Lagrangian drift or Lagrangian mean flow.

[14] There are reasons to expect this correlation to be much less than 1.0. If the buoy drift is dominated by surface currents rather than windage, those currents may not be wind-driven, e.g., tidal currents, boundary currents. Even in cases where wind-driven current dominates, some offset in direction can be expected, associated with Coriolis. For example, the idealized Ekman surface current is, according to theory, offset from the wind direction by 45°, and Nansen observed an offset of 20° to 40° (Pond and Pickard 1983, Chapter 9) at a high latitude. Moreover, in veering winds or downwind of a coastline, the windsea direction may be offset somewhat from the instantaneous wind direction.

Further study is needed to quantify this effect. If we find that the impact is significant, we should we make some effort to put all buoy spectra into a common frame of reference $E(\omega)$ or $E(\sigma)$. This is not a simple task. The buoy provides the spectral density and four Fourier coefficients describing the directional distribution (Longuet-Higgins et al. 1963, Steele et al. 1985): $E(f)$, $a_1(f)$, $b_1(f)$, $a_2(f)$, $b_2(f)$. Since Doppler shift is directionally dependent, we expect that a directional estimator such as Maximum Entropy Method would need to be applied to transform the buoy observations into directional spectra $E(\sigma,\theta)$ prior to applying the Doppler shift. A second problem is that Doppler shift of $E(f_{ob},\theta)$ to $E(\sigma,\theta)$ requires a remapping in frequency space; otherwise, each observation will have its own frequency grid. A third problem is that an estimate of the surface current is needed to transform to $E(\sigma)$. An ocean model can, of course, be used, but errors are inevitable.

Janssen (2009) discusses a phenomenon called "Stokes-frequency correction". This is a "Doppler shift of the frequency by nonlinear effects". In deep water, this results in a shift toward higher frequencies. It is associated with nonlinearity while the discussion above is not. Also, it affects the waves themselves and would not necessarily affect one platform (say, the moored buoys) more than another. In other words, it does not appear to help with explaining our discrepancies. Nevertheless, we find it credible that it has significant impact on measurement of high frequency waves and so it deserves further consideration.

### 6.2. Buoy response, motion sensor, and onboard processing

The ability of a buoy to follow the motion of the ocean surface is always a concern, particularly for cases where the wavelength is not large relative to the buoy diameter, i.e., large buoys have problems measuring short waves. The Response Amplitude Operator (RAO) quantifies the responsiveness of the buoy: an RAO equal to 1.00 is ideal, where values less than unity indicate a damped response, and values larger than unity indicate resonant excitation. In cases where the RAO deviates from unity, a correction can be applied during onboard processing, but only if the RAO is well understood. Note that even when the RAO is unity, the buoy data can still be erroneous if a RAO correction is incorrectly applied.

In the comparison of four buoy types in Figure 18, the DWR is strongly inconsistent with the other three. However, buoy size would not lead us to expect this, since it is 0.9 m diameter, while the NDBC foam hulls are 3 m diameter, being large enough to provide a stable platform for a variety of meteorological instruments. Insofar as our results indicate consistency between NDBC and the two small drifting buoy types, this may indicate that the NDBC buoys use a "good" RAO correction or non-correction, but if Doppler effect is significant, it complicates this hypothesis.

In Section 4.2, we use 0.4 Hz for our comparisons. At this frequency, Ribe (1982) shows the Datawell Waverider (DWR) RAO near 1.00, meaning that the RAO should not be expected to cause a problem. Ribe (1982) shows the DWR RAO deviating from 1.00 after around 0.48 to 0.50 Hz, and state that no correction is applied at the high frequencies. James Behrens (CORDC/CDIP), in a personal communication, confirmed that no correction is applied. Based on this information alone, we should not be concerned about the accuracy of the DWR observations at 0.4 Hz.

In Rogers (2017), we evaluated the performance of an NDBC 3-m discuss buoy against the OSP DWR buoy, using comparisons of mean square slope (MSS) against model output. We also presented a direct comparison by USACE (provided by Dr. Robert Jensen) between the two buoy types (NDBC 46042 vs. NDBC 46114 / CDIP 185) at Monterey Bay, CA. These comparisons indicated a strong negative bias of the 3-m buoy relative to the DWR buoy in the high frequencies. At the time, we were under the impression that the DWR buoy is the "gold standard" for buoy measurement (e.g., O'Reilly et al. (1996), in context of directional measurements of low frequencies), and so we concluded that the problem was with the larger buoy: a damped response and/or poor RAO correction for the 3-m NDBC buoy. The present report suggests that this conclusion by Rogers (2017) was flawed: suspicion should have been applied also to the smaller DWR buoy. Interestingly, the evaluation of MSS in Rogers (2017) indicated a strong consistency between the 3-m NDBC buoy and the drifting UCSD/SIO/CORDC buoys. In fact, we can find no aspect of the analysis presented in Rogers (2017) that is inconsistent with those in the present report. The only difference is in the conclusions drawn from those results.

Collins et al. (2024), in their Figure 15, show that the slope of the non-directional spectrum from a DWR buoy is smaller than that from two other instruments (and Acoustic Doppler Current Profiler, and a Sofar Spotter buoy), and deviates from the expected $f^{-4}$ slope after 0.4 Hz. They state: "*The departure of the high-frequencies from the theoretical expectation suggests that the 17-m Waverider is misrepresenting the wave energy from 0.40–0.60 Hz. This could be related to Datawell Waverider's hydrodynamic response function, which has a peak around 0.80 Hz (Ribe 1982; Thomson et al. 2015). This implies a quasi-resonant "bobbing" motion for 0.4 > f >1 Hz during low wave conditions.*" Their Figure 15 shows the dependence on wave height and frequency on log scale, where our Figure 18 shows the dependence on wind speed in linear scale, but to the extent that we can scrutinize their plot around 0.35 to 0.45 Hz, it is not inconsistent with the 0.06 vs. 0.08 $m^{-+}$/Hz discrepancy in our Figure 18.

The discussion of Collins et al. (2024) continues: *However, disagreement of similar magnitude but starting at even lower frequencies was found between a strapped-down IMU and HIPPY system on the same 3-m discus buoy (see Figure 9 and A2 of Collins and Jensen (2022)). Thus, this error may somehow be related to the performance of the HIPPY system."* In other words, the issue may be with the sensor on the buoy, rather than the RAO of the buoy.

Noise, and filters used to suppress noise, may also lead to systematic bias in the tail levels. However, this is more of a problem for low frequencies, and we do not expect this to be a significant issue with the comparisons herein.

If the wave measurement community were to conclude that the DWR buoys overestimate tail levels, a post-facto correction that depends on wind speed and frequency could be designed using comparisons similar to our Figure 18. This would allow us to continue using the dataset for model evaluation. However, we should keep in mind that our Figure 18 are the bin-averaged, implying that this could be an overly blunt correction. Collins (personal communication) suggested an improvement on this concept, which is to incorporate the buoy's check factor[15] into the correction. If the check factor correlates with the error, this would provide a more precise correction.

[15] This is the ratio of horizontal displacements to vertical displacements, see https://cdip.ucsd.edu/m/documents/data_processing.html#buoy-check-factors .

## 7. Summary of conclusions and recommendations

### 7.1. Primary conclusions

We evaluate the energy level in the high frequencies of the ocean wave spectra from the following wave buoy types:

1) UCSD/SIO/CORDC miniature wave buoys
2) Sofar Spotter minature wave buoys
3) NOAA/NDBC buoys, primarily 3-meter foam buoys with SCOOP payload
4) Datawell Waverider (DWR) buoys operated by UCSD/SIO/CDIP.

Our primary conclusions are as follows:

- We find that (1)-(3) are, in the mean, consistent, while (4) reports significantly higher energy levels.
- Our calibrations of the wave models SWAN and WW3, used operationally by the USN, agree better with (4).
- Inspection of tail slope from (4) also suggest a problem with the DWR buoys at high frequencies, but the problems are not obvious and are subjective. Further, (1)-(3) were not subjected to the same scrutiny in this report.
- Model-data comparisons of $H_{m0B4}$ at Ocean Station Papa vs. the DWR (type 4) buoy show very high correlation (CC=0.97), while correlation is lower against the CORDC (type 1, CC=0.89) and Sofar (type 2, CC=0.95) buoys. Dissimilar local wave climate may have a role in this outcome, but it is nevertheless an encouraging sign for the future use of the DWR buoys in evaluation of high frequency wave energy, as correlation is not affected by systematic bias. If systematic bias exists with (4), post facto corrections are possible.
- Though most of our analysis of (4) is based on one buoy (the OSP buoy), we confirmed that this behavior is a feature common to buoy type (4) (Figure 15).
- The dependence of high-frequency energy levels on wind speed is a useful method for identification of sensor or RAO discrepancies (Figure 18) and time-dependent malfunctions such as biofouling (Appendix B).

### 7.2. Secondary conclusions

Our secondary conclusions are as follows:

- The primary conclusions drawn from model-data comparisons are insensitive to model forcing, model version, or model settings.
- We find a non-monotonic trend for $H_{m0B4}$ observed by Sofar buoys vs. wind speed. This is consistent with prior evaluations of the dependency of drag coefficient, wind stress, and mean square slope with wind speed. A remarkable feature of the $H_{m0B4}$ vs. wind speed comparison however is that the peak occurs around $U_{10N}$=15 m/s, which is a lower wind speed than was reported for the other parameters in those prior studies. The wave model applied here does not reproduce the non-monotonic trend.

- Based on our reading of Ribe (1982) and information and expert opinions provided by Dr. Collins (USACE), we feel that—if the DWR buoys are overestimating high frequency energy levels—the problem is probably not associated with the hull response (RAO), but rather is caused by some other issue, such as the sensor.

### 7.3. Recommendations

Our recommendations are as follows:

- Analysis should be performed to quantify the likely effect of Doppler shift on the tail levels of the drifting buoys.
- If future analysis indicates that (1)-(3) are accurate, and (4) is biased high, then
    - our models will need to be recalibrated, and
    - post-facto corrections to the DWR spectra be made. This can be done (crudely) on a "per frequency" basis, or better yet, as suggested by Dr. Collins, the buoy check factor may also be used to provide a more precise correction.
- We recommend continued use of DWR/CDIP spectra, even at high frequencies, especially considering the strong model-buoy correlations described above.
- Though we are not yet convinced that is the dominant cause of discrepancies between buoy types used here, we believe it likely that Doppler shift significantly affects validations of high frequency energy of ocean wave models, and we recommend that efforts be made to account for this.

### Acknowledgments

This work was funded by the Office of Naval Research through the NRL Core Program, Program Element 0602435N. During FY25, it was funded by the NRL Core 6.2 project "Coupled Prediction of Ocean Waves at Extended Ranges (C-POWER)", PI M. Janiga, Work Unit 62A1C5 and NRL Core 6.2 project, "Developing a Spectral Wave Data Assimilation System for WAVEWATCH III", PIs M. Orzech and H. Ngodock, Work Unit 996D47. During FY26, it was funded by NRL Core 6.2 project, PI E. Rogers, Work Unit 62A1P9. It is approved for public release.

We are grateful that all observational data used in this study are provided freely to the research community, and acknowledge the significant effort and resources required to make this available. We thank Eric Terrill, Mark Otero, Peter Rogowski, and others at CORDC for the UCSD/SIO/CORDC mini-buoy dataset. We thank APL/UW (Jim Thomson and others) for the OSP wave buoy dataset. We thank Pieter Smit and others at Sofar Oceans for the Sofar Spotter dataset. We thank the Ocean Climate Stations (OCS) Project Office of NOAA/PMEL (Meghan Cronin and others) for the OSP meteorological dataset. We thank James Behrens and others at CDIP for the UCSD/SIO/CDIP datasets. We thank NOAA/NDBC for data from their moored buoy network.

We thank ECMWF and the European Commission for sharing their ERA5 wind and ice products through C3S of the Copernicus Earth Observation Programme.

We thank Drs. Tripp Collins (USACE), David Wang (NRL), Jie Yu (NRL), Eric Terrill (CORDC), and Sophia Merrifield (CORDC) for valuable insights, information, and discussion.

Dr. Collins provided feedback on earlier versions of this report which have been used during revisions.

## Appendix A. Impact of diagnostic tail cutoff frequency in WW3

In Section 2.3.1, we implied that the FXFM3 parameter in WW3 has a significant impact on the $H_{m0B4}$ and $E(0.4\ \text{Hz})$ results presented in this report. We elaborate further in this appendix. The WW3 simulations used here are taken from simulations using 0.28° resolution NAVGEM forcing, with valid output from 08 November 2015 to 1 February 2016, run four times with different values of FXFM3. The $H_{m0B4}$ values with $f_1$=0.29 Hz and $f_2$=0.58 Hz from these models are compared against the same parameter from OSP buoy data from January 2016 and 2020. The setting of FXFM3=10 is similar to the default setting in Ardhuin et al. (2010), FXFM3=9.9 and causes more of the high frequency tail to be in the prognostic range (computed from source terms). The setting of FXFM3=2.5 matches the default setting in WW3 v6 and v7 and causes more of the high frequency tail to be in the diagnostic range (computed by application of diagnostic tail attached at the highest prognostic frequency). By treating more of the spectrum as diagnostic, the tail levels better match those of the OSP buoy.

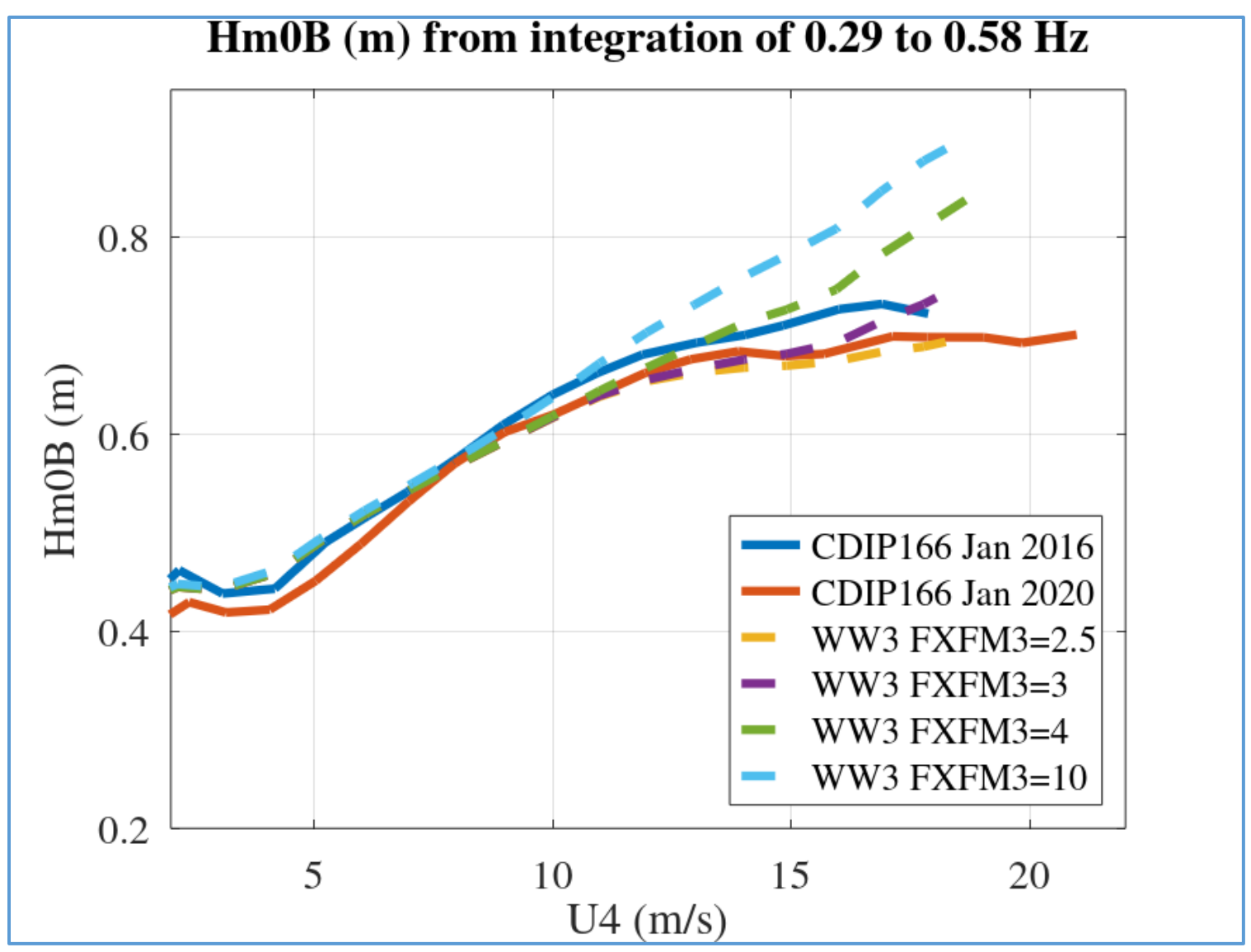

Figure 21. $H_{m0B4}$ as a function of wind speed (bin-averaged values), comparing WW3 output with different FXFM3 settings against OSP buoy data.

## Appendix B. Biofouling and the OSP buoy (CDIP 166)

Deployments of the APL/UW/CDIP Ocean Station Papa (OSP) buoy (CDIP 166) are tabulated in Table I. In Figure 22 and Figure 23, we compare $H_{m0B4}$ as a function of wind speed (bin-averaged values), where each curve is based on a single month of the deployment. The left panel of Figure 22 corresponds to months near the beginnings of the first eight deployments. The right panel of Figure 22 corresponds to months near the ends of six of the first eight[16] deployments. Figure 22 indicates that the deployments 2 and 3 were operating normally at the beginning of the deployments and afflicted by biofouling at the ends of the deployments. The biofouling during deployment #2 is documented by Thomson et al. (2015). The biofouling is associated with the paint used (reference for deployment #2: Thomson et al. (2015); reference for deployment #3: J. Thomson personal communication, 24 February 2020). Figure 23 corresponds to months during deployment #3, which is the second of two deployments afflicted by biofouling. This indicates that the biofouling became a problem during June or July 2016.

Table I. Information valid 5 November 2025, taken from the CDIP webpage, https://cdip.ucsd.edu/m/deployment/?stn=166, showing deployments of the OSP buoy. The deployments afflicted by biofouling are indicated with gray shading.

| Deployment # | Start | End |
|---|---|---|
| 1 | 16 June 2010 | 4 October 2012 |
| 2 | 4 October 2012 | 1 January 2015 |
| 3 | 1 January 2015 | 5 October 2017 |
| 4 | 28 July 2018 | 14 December 2019 |
| 5 | 14 December 2019 | 6 May 2020 |
| 6 | 16 August 2020 | 25 April 2021 |
| 7 | 25 April 2021 | 1 January 2022* |
| 8 | 16 May 2022 | 17 May 2023 |
| 9 | 17 May 2023 | 25 June 2023 |
| 10 | 02 June 2024 | 23 May 2025 |
| 11 | 23 May 2025 | Active |

* The end of deployment is given as 16 May 2022 on the CDIP website above, but we find that the last date with valid data is 1 January 2022.

[16] Two are not included, since wind observations are missing and incomplete for March 2021 and April 2023, respectively.

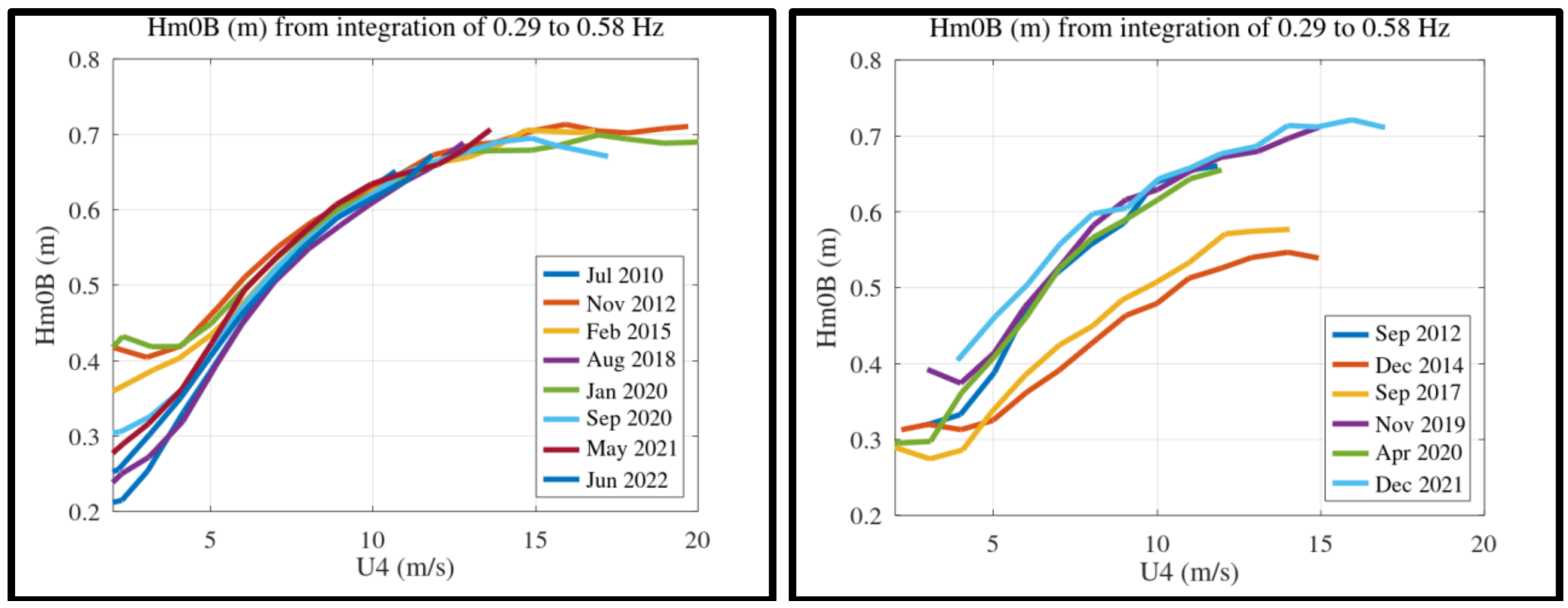

Figure 22. $H_{m0B4}$ as a function of wind speed (bin-averaged values), from OSP buoy data at different months. Left panel: months at the beginning of eight deployments. Right panel: months at the end of six of the first eight deployments. Wind speed is taken from the NOAA/PMEL buoy at OSP.

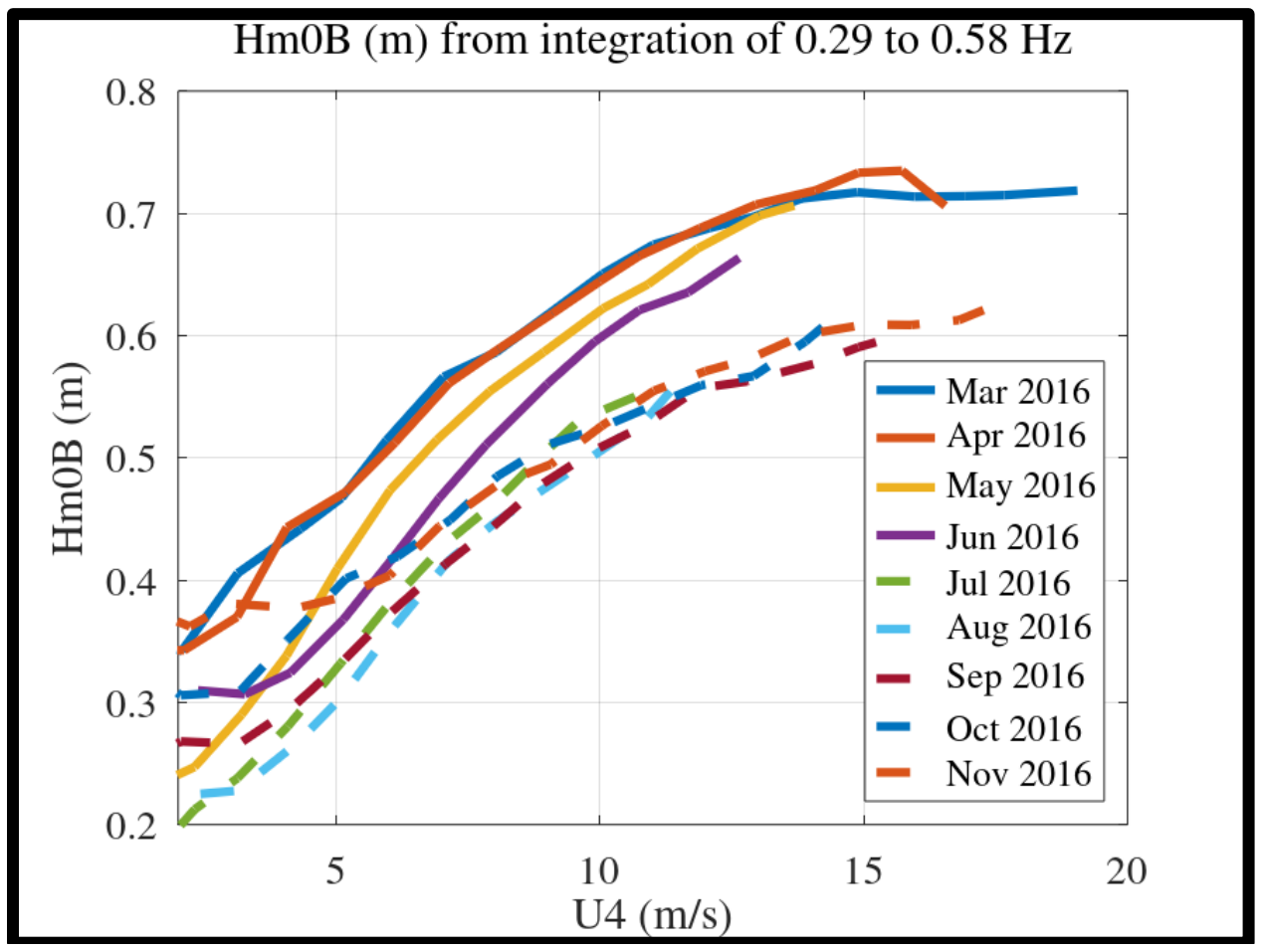

Figure 23. $H_{m0B4}$ as a function of wind speed (bin-averaged values), from OSP buoy data at different months during the second of two deployments afflicted by biofouling, January 2015 to October 2017.

## Appendix C. Additional validations using Sofar dataset, February 2022

In Section 3.4, we presented brief comparisons of $H_{m0B4}$, WW3 vs. Sofar measurements, for February 2022. In this appendix, we present additional, related comparisons. First, we present results for other frequency bands, omitted from Section 3.4 since our focus is on high frequencies (Figure 24 and Figure 25). Second, we evaluate the sensitivity of our reported results to the values of $f_1$ and $f_2$ used to define the upper and lower bounds of the $H_{m0B4}$ calculation (Figure 27 and Figure 26).

Figure 24 extends the comparison of Figure 6 to show the spatial distribution of bias for total wave height $H_{m0}$ and three additional $H_{m0B}$ bands ($H_{m0B1}$ for low frequencies, $H_{m0B2}$ for mid-to-low frequencies, and $H_{m0B3}$ for mid-to-high frequencies). This illustrates how the systematic positive bias in WW3's $H_{m0B4}$ is not seen in the other bands. Figure 25 extends the comparison of Figure 7 to show the overall scatter plots for $H_{m0}$ and the three additional $H_{m0B}$ bands. Again, this illustrates how the systematic positive bias in WW3's $H_{m0B4}$ (+9%) is not seen in $H_{m0}$ or the other $H_{m0B}$ bands (+1%, -6%, +1%, +1%).

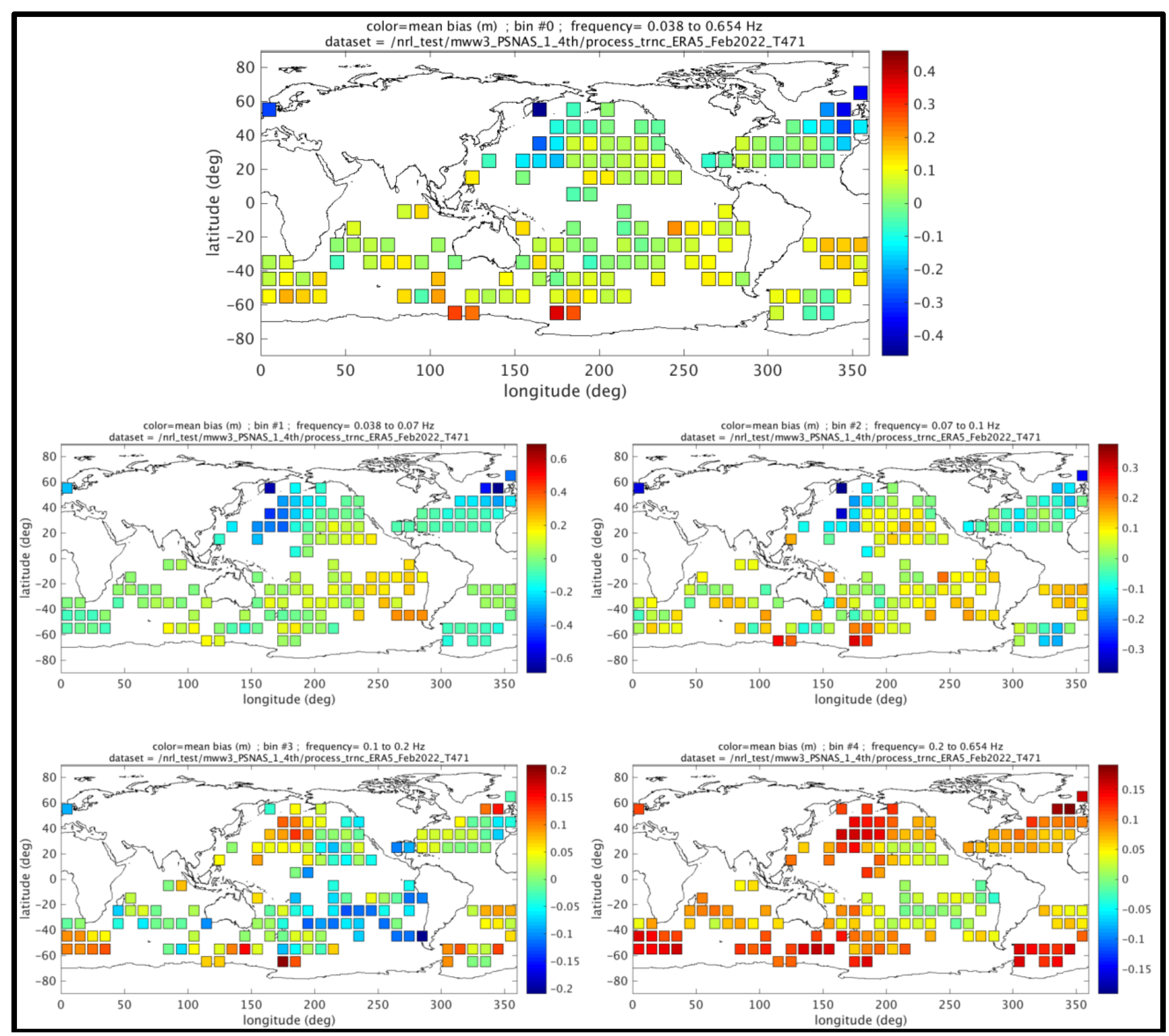

Figure 24. Spatial distribution of the bias of WW3 relative to the Sofar buoys for February 2022. Here, WW3 is forced by ERA5. Top: $H_{m0}$. Center left: $H_{m0B1}$. Center right: $H_{m0B2}$. Lower left: $H_{m0B3}$. Lower right: $H_{m0B4}$. The lower right panel ($H_{m0B4}$) is identical to the lower panel of Figure 6.

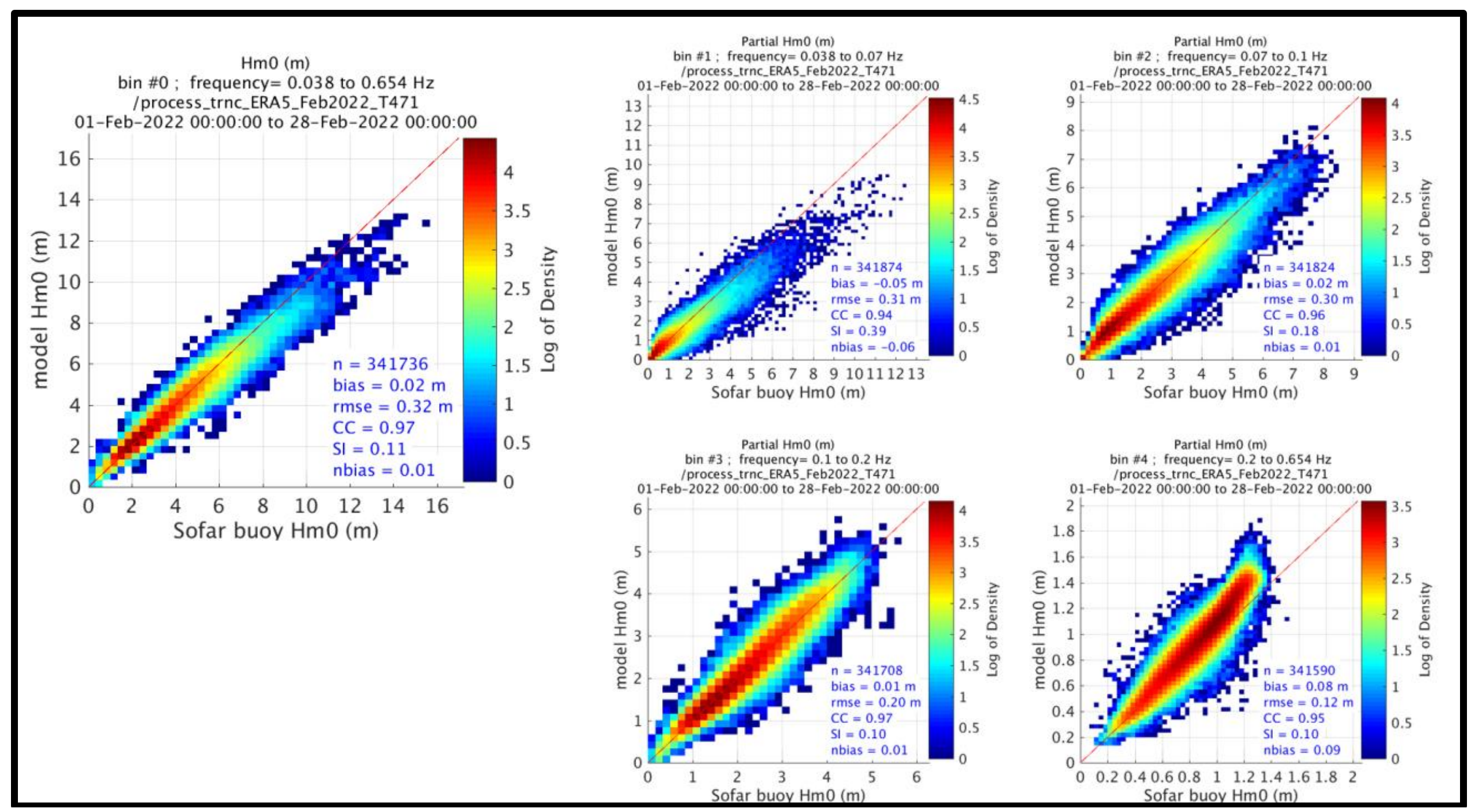

Figure 25. Scatter plots of $H_{m0}$ and $H_{m0B}$ of WW3 hindcast relative to Sofar buoys for the month of February 2022. The WW3 hindcast is forced by ERA5 fields. Left: $H_{m0}$. Center top: $H_{m0B1}$. Top right: $H_{m0B2}$. Center bottom: $H_{m0B3}$. Bottom right: $H_{m0B4}$. The bottom right panel ($H_{m0B4}$) is identical to the right panel of Figure 7.

Figure 27 illustrates the sensitivity of spatial distribution of model bias to the values of $f_1$ and $f_2$ used to define the upper and lower bounds of the $H_{m0B4}$ calculation. The values of $f_1$ and $f_2$ are indicated in the captions. We find that these definitions affect the outcome, but do not affect the broad conclusions that one might draw from the diagrams.

Figure 26 illustrates the sensitivity of the overall scatter plots to the values of $f_1$ and $f_2$. Again, the outcome is noticeably different, but not in a way that would affect conclusions drawn. The outcome is predictable: when the $H_{m0B4}$ is defined over a narrower band ($f_1 = 0.29$ Hz and $f_2$ =0.58 Hz), the median $H_{m0B4}$ decreases for both model and observations, model bias decreases, and normalized model bias increases.

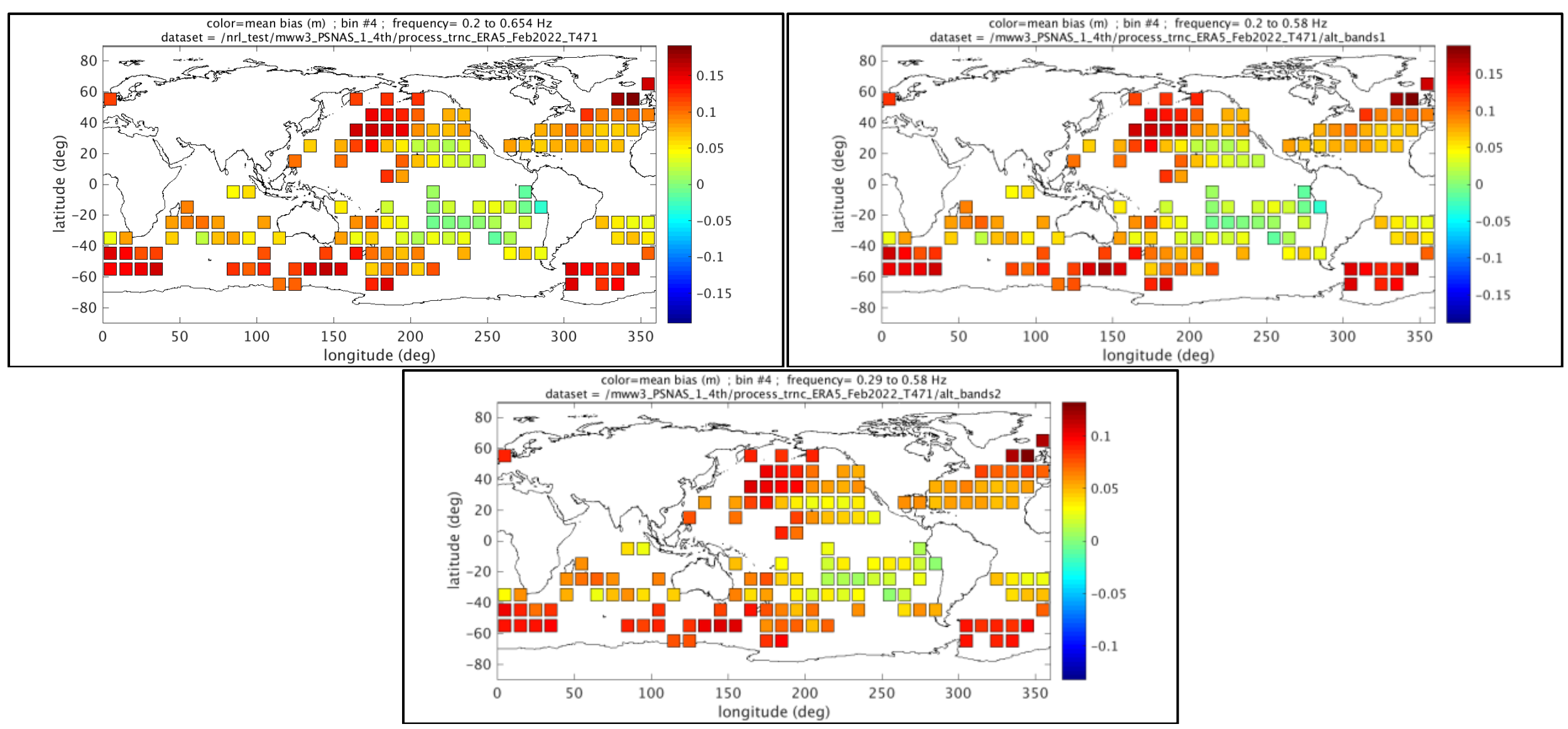

Figure 26. Spatial distribution of the bias of WW3 relative to the Sofar buoys for February 2022. The WW3 hindcast is forced by ERA5 fields. Upper left panel: $f_1 = 0.2$ Hz and $f_2 =0.654$ Hz. Upper right panel: $f_1 = 0.2$ Hz and $f_2 =0.58$ Hz. Lower panel: $f_1 = 0.29$ Hz and $f_2 =0.58$ Hz.

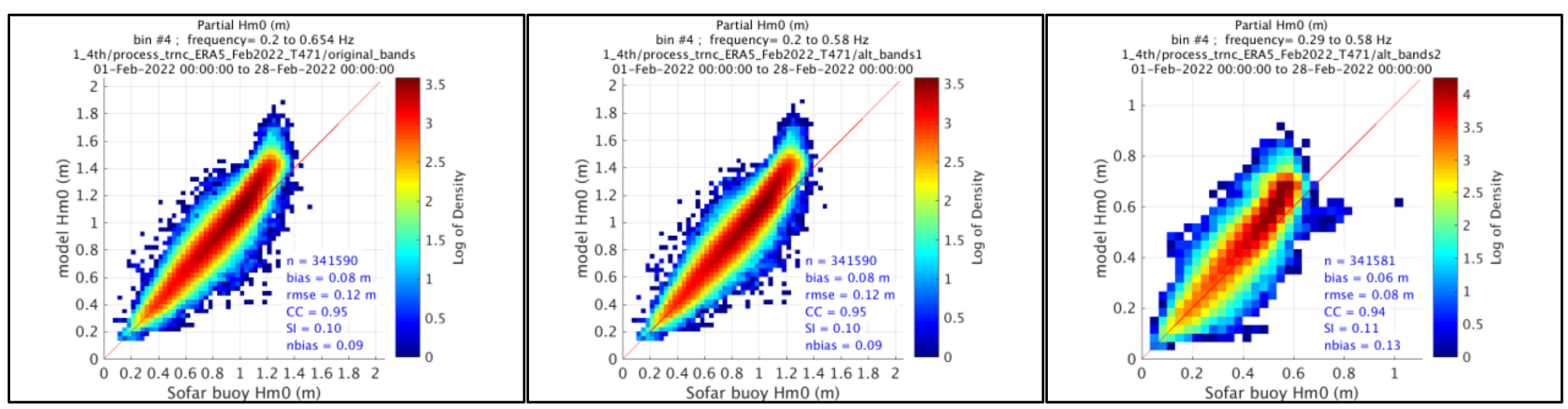

Figure 27. Scatter plots of $H_{m0B4}$ for a WW3 hindcast, relative to Sofar buoys for the month of February 2022. The WW3 hindcast is forced by ERA5 fields. Left panel: $f_1 = 0.2$ Hz and $f_2 =0.654$ Hz. Center panel: $f_1 = 0.2$ Hz and $f_2 =0.58$ Hz. Right panel: $f_1 = 0.29$ Hz and $f_2 =0.58$ Hz. The left panel is identical to the right panel of Figure 7 and the bottom right panel of Figure 25.

## Appendix D. Parametric spectrum

The parametric spectrum used in figures herein is based on multiple sources. The basic form is from Toba (1973), later used by Forristall (1981), Battjes et al. (1987), Toba (1988), Holthuijsen (2007), and many others.[17]

We adopt the notation of Battjes et al. (1987), $E(f) = \beta g u_* (2\pi)^{-3} f^{-4}$, and use $\beta = 0.096$, which is given by Holthuijsen (2007) as "an average of [Battjes et al. 1987] Table 2 field observations". The formula is not valid when the frequency is below the lower limit of the equilibrium range, $f < f_{eq}$. For this lower limit, we use $f_{eq} = 1.5 f_p$ , as given in Tsagareli (2009) and Vincent et al. (2019). For the peak frequency $f_p$, we use the fully-developed spectrum of Pierson and Moskowitz (1964), $f_p = 0.13 g / U_{10}$. Above the upper limit of the equilibrium range, following Tsagareli (2009) and Vincent et al. (2019), we attach a tail with slope $f^{-5}$ at the transition frequency $f_t$, and so $E(f) = \beta g u_* (2\pi)^{-3} f_t f^{-5}$ at those frequencies. For $f_t$, we use the relation from Forristall (1981): $f_t = 0.29 / u_*$. [Other formulae for $f_t$ can be found in the literature: Kahma and Calkoen (1992) and Lenain and Melville (2017).] For estimation of $u_*$, we use the method of Hwang (2011). His equation 5 is $C_{10} = 10^{-5}(-0.160 U_{10}^2 + 9.67 U_{10} + 80.58)$, valid for $U_{10} \leq 50$ m/s, and $u_* = U_{10}\sqrt{C_{10}}$. By combining the above, we have a parametric formula for spectral density that depends only on frequency and wind speed, $E(f, U_{10})$.

[17] On notation: The Toba (1973) formula for equilibrium spectra is $\phi_g = \alpha_g g u_* \sigma^{-4}$ (his equation 2.12). This is rendered in Forristall (1981) as $S(f) = \alpha_4 g u_* f^{-4}$ (his equations 8, 9, and 12) ; in Battjes et al. (1987) as $E(f) = (2\pi)^{-3} \beta g u_* f^{-4}$ (their equation 3); in Toba (1988) formula as $\phi(\sigma) = \alpha_s g u_* \sigma^{-4}$ (his equations 5 and 9); in Holthuijsen (2007) as $E(\omega) = \alpha_{Toba} g u_* \omega^{-4}$ (his Note 6C). These are all consistent, though the coefficients of proportionality are not interchangeable. Vincent et al. (2019) use an inconsistent form, where $u_*$ is replaced by the 10-meter wind speed (their equations 1 and 8).